\documentstyle[12pt]{article}

\newcommand{\ba}{\begin{eqnarray}}
\newcommand{\ea}{\end{eqnarray}}
\begin{document}
\pagestyle{plain}

\title{Electromagnetic Form Factors in a Collective Model of the Nucleon}
\author{R.~Bijker$^{a}$, F.~Iachello$^{b}$ and A.~Leviatan$^{c}$\\
\and
\begin{tabular}{rl}
$^{a}$&Instituto de Ciencias Nucleares, U.N.A.M.,\\
      &A.P. 70-543, 04510 M\'exico D.F., M\'exico\\
$^{b}$&Center for Theoretical Physics, Sloane Laboratory,\\
      &Yale University, New Haven, CT 06520-8120, U.S.A.\\
$^{c}$&Racah Institute of Physics, The Hebrew University,\\
      &Jerusalem 91904, Israel
\end{tabular}}
\date{}
\maketitle
\noindent
\vspace{6pt}
\begin{abstract}
We study the electromagnetic form factors of the nucleon in a
collective model of baryons. Using the algebraic approach to hadron
structure, we derive closed expressions for both elastic and transition
form factors, and consequently for the helicity amplitudes that
can be measured in electro- and photoproduction.
Effects of spin-flavor symmetry breaking and of swelling of hadrons with
increasing excitation energy are considered.
\end{abstract}
\begin{center}
\end{center}

\newpage
\section{Introduction}

Form factors are an important ingredient in understanding the structure of
hadrons. Elastic form factors of the nucleon have been measured several
times \cite{bosted92} up to relatively large momentum transfer,
$Q^2\approx 20$ (GeV/c)$^2$. In the absence of detailed solutions of QCD in
the nonperturbative regime, they have been described by models.
Traditionally, Vector Dominance Models \cite{vmd} have been used to fit the
data in the low $Q^2$ region.
For $Q^2 \gg M^2$, where $M$ is the nucleon mass, perturbative
QCD has been used \cite{pQCD}. Other approaches include constituent quark
models \cite{qmff}, QCD sum rules \cite{qcdsum} and quark-diquark models
\cite{qdiq}. Inelastic (transition) form factors have also been measured
\cite{burkert92}, although not as accurately as the elastic ones.
A remeasurement of these form factors will form an important
part of the experimental programs at various facilities, {\it e.g.}
CEBAF ($N^{*}$ collaboration) and MAMI. Extensive calculations
have been carried out in the nonrelativistic and relativized quark
models \cite{ki,warns,closeli,capstick}.

In this article, we present another method which can describe
simultaneously both elastic and inelastic form factors. This method
is semi-phenomenological, in the sense that it assumes a certain form for the
elastic form factors, and then calculates all other form factors by making use
of the algebraic approach to hadron structure \cite{bil}.
The main aspect of the paper is the presentation of results for form factors
and helicity amplitudes in an explicit analytic form that allows one to study
models of hadron structure having  the same spin-flavor structure. In addition,
we investigate two additional aspects of the nucleon form factors, arising from
breaking of the ``effective'' spin-flavor symmetry in the three constituent
channel, and of swelling of hadrons with increasing excitation energy. We find
that, even if we attribute the entire neutron electric form factor to breaking
of $SU_{sf}(6)$, this breaking still does not significantly affect other
observable quantities, while the stretching of hadrons with increasing
excitation energy plays a significant role.  The phenomenological breaking
needed to describe $G^{n}_{E}$ is much too large when compared with QCD flavor
breaking mechanisms \cite{cmt}, and it worsens the description of 
$G^{p}_{E}$. We conclude therefore, as other authors do, that meson cloud
corrections play an important role in $G^{n}_{E}$.

\section{Collective Model of Baryons}
\setcounter{equation}{0}

We begin by reviewing the algebraic approach to baryon structure \cite{bil}.
This approach can be used for any constituent model, but we consider in
this article a collective (string-like) model with the configuration
depicted in Fig.~1. The relevant degrees of freedom of this
configuration are the two Jacobi coordinates
\ba
\vec{\rho} &=& \frac{1}{\sqrt{2}}
(\vec{r}_1 - \vec{r}_2) ~,
\nonumber\\
\vec{\lambda} &=& \frac{1}{\sqrt{6}}
(\vec{r}_1 + \vec{r}_2 - 2\vec{r}_3) ~, \label{jacobi}
\ea
where $\vec{r}_1$, $\vec{r}_2$ and $\vec{r}_3$ denote the end points of
the string configuration.
In the algebraic approach, the Jacobi coordinates, $\vec{\rho}$ and
$\vec{\lambda}$, and their conjugate momenta, $\vec{p}_{\rho}$ and
$\vec{p}_{\lambda}$, are quantized (up to a canonical transformation)
with boson operators
\ba
b^{\dagger}_{\rho,m} &=&
\frac{1}{\sqrt{2}} ( \rho_m - i \, p_{\rho,m} ) ~,
\;\;\;
b_{\rho,m} =
\frac{1}{\sqrt{2}} ( \rho_m + i \, p_{\rho,m} ) ~,
\nonumber\\
b^{\dagger}_{\lambda,m} &=&
\frac{1}{\sqrt{2}} ( \lambda_m - i \, p_{\lambda,m} ) ~,
\;\;\;
b_{\lambda,m} =
\frac{1}{\sqrt{2}} ( \lambda_m + i \, p_{\lambda,m} ) ~, \label{boson}
\ea
with $m=-1,0,1$, and an additional scalar boson, $s^{\dagger}$,
$s$ is introduced. These operators satisfy usual boson commutation
relations and operators of different type commute.
Their number-conserving bilinear products
generate the Lie algebra of $U(7)$
whose elements serve in the expansion of physical operators (the mass operator
and transition operators). The $U(7)$ algebra enlarges the $U(6)$ algebra of
the harmonic oscillator quark model \cite{hoqm},
still describing the dynamics of two vectors.
The $s$-boson does not
introduce a new degree of freedom, since for a given total boson number $N$
it can always be eliminated
$s\rightarrow \sqrt{N-\hat n_{\rho}-\hat n_{\lambda}}$ (Holstein-Primakoff
realization of $U(7)$). Its introduction is
just an elegant and efficient way by means of which the full dynamics of two
vectors can be investigated, including those situations in which there is
strong mixing of the oscillator basis (collective models).
For a system of interacting bosons all states of the model space are
assigned to the totally symmetric representation $[N]$ of
$U(7)$. This representation contains all oscillator shells with
$n=n_{\rho}+n_{\lambda}=0,1,2,\ldots, N$. The value of $N$ determines
the size of the model space and, in view of confinement, is expected to be
large. The geometric structure of baryons is thus described by the algebra
of
\ba
{\cal G}_r \equiv U(7) ~.
\ea
The full algebraic structure is obtained by combining this part with
the internal spin-flavor-color part
\ba
{\cal G}_i \equiv SU_{sf}(6) \otimes SU_{c}(3) ~.
\ea

As discussed in detail in Ref. \cite{bil},
the object of Fig.~1 is a top. If the three strings have
equal length and equal relative angles, the top is oblate and has $D_{3h}$
point group symmetry. The classification under $D_{3h}$ is equivalent to
the classification under permutations and parity \cite{bunker}.
States are characterized by
$(v_1,v_2);K,L_t^{P}$, where $(v_1,v_2)$ denote the vibrations
(stretching, bending); $K$ denotes the projection of the rotational
angular momentum $L$ on the body-fixed symmetry axis, $P$ the parity and
$t$ the symmetry type of the state under $D_3$ (a subgroup of $D_{3h}$),
or equivalently the symmetry
type under $S_3$, the group of permutations of the three end points
($S_3$ and $D_3$ are isomorphic). Both groups have one-dimensional symmetric
and antisymmetric representations and a two-dimensional representation,
called $S$, $A$, $M$ for $S_3$ and $A_1$, $A_2$, $E$ for $D_3$, respectively.
The notation in terms of $S_3$ is the one used in constituent quark
models \cite{nrqm,rqm}. The permutation symmetry of the geometric
part must be the same as the permutation symmetry of the spin-flavor part
in order to have total wave functions that are antisymmetric (the color part
is a color singlet, {\it i.e.} antisymmetric). Therefore one can also use
the dimension of the $SU_{sf}(6)$ representations to label the states:
$S \leftrightarrow A_1 \leftrightarrow 56$,
$A \leftrightarrow A_2 \leftrightarrow 20$,
$M \leftrightarrow E \leftrightarrow 70$.
In Ref. \cite{bil}, a $S_3$-invariant mass operator was used, consisting of
spatial and spin-flavor contributions, to obtain a description of the mass
spectrum of nonstrange baryons with a r.m.s. deviation of 39 MeV.
The nonstrange baryon resonances were identified
with rotations and vibrations of the string.
The corresponding wave functions, when expressed in a harmonic oscillator
basis, are spread over many shells and hence are truly collective.

\section{Electromagnetic Form Factors}
\setcounter{equation}{0}

Electromagnetic form factors appear in the coupling of baryons with the
electromagnetic field. In constituent models, the (point-like) constituent
parts are coupled to the field \cite{copley}. The current is then
reduced \cite{warns,closeli} to a nonrelativistic part,
a spin-orbit part and a non-additive part associated with
Wigner rotations and higher order corrections. We discuss here the
nonrelativistic contribution to the form factors for nonstrange baryons.
Transverse, longitudinal and scalar couplings can be expressed in terms
of the operators \cite{bil}
\ba
\hat U &=& \mbox{e}^{ -ik \sqrt{\frac{2}{3}} \lambda_z} ~,
\nonumber\\
\hat T_{m} &=& \frac{i m_{3} k_0}{2} \left(
\sqrt{\frac{2}{3}} \, \lambda_m \,
\mbox{e}^{ -ik \sqrt{\frac{2}{3}} \lambda_z} \, + \,
\mbox{e}^{ -ik \sqrt{\frac{2}{3}} \lambda_z} \,
\sqrt{\frac{2}{3}} \, \lambda_m \right) ~, \label{ut}
\ea
(with $m=\pm 1,0$) which act on the spatial part of the baryon wave function.
Here $\vec{k}=k \hat z$ is the photon momentum, $k_0$ the
photon energy, and $m_{3}$ the mass of the third constituent.
The form factors of interest in photo- and
electroproduction as well as in elastic electron scattering are proportional
to the matrix elements of these operators between initial and final states.
These matrix elements can be evaluated in coordinate or momentum space as done
in the nonrelativistic \cite{copley,ki} or relativized quark model
\cite{rqm,warns,closeli,capstick}. Following Ref.~\cite{bil} we prefer to
use an algebraic method to evaluate the matrix elements of Eq.~(\ref{ut}).
In order to do this, we first express the operators in
Eq.~(\ref{ut}) in terms of generators of the algebra of $U(7)$
\ba
\hat U &=& \mbox{e}^{ -i k \beta \hat D_{\lambda,z}/X_D } ~,
\nonumber\\
\hat T_m &=& \frac{i m_{3} k_0 \beta}{2 X_D} \left( \hat D_{\lambda,m} \,
\mbox{e}^{ -i k \beta \hat D_{\lambda,z}/X_D } \, + \,
\mbox{e}^{ -i k \beta \hat D_{\lambda,z}/X_D } \, \hat D_{\lambda,m}
\right) ~. \label{emop}
\ea
Here the dipole operator
$\hat D_{\lambda,m} = (b^{\dagger}_{\lambda} \times s -
s^{\dagger} \times \tilde b_{\lambda})^{(1)}_m$,
with $\tilde b_{\lambda,m} = (-1)^{1-m}b_{\lambda,-m}$,
is a generator of $U(7)$ which transforms as a vector ($L^P=1^{-}$)
under rotations, is even under time-reversal and has the same character
under permutations as the Jacobi coordinate $\lambda_m$.
The coefficient $X_D$ is a normalization factor given by the reduced
matrix element
$X_D = |\langle 1^-_M || \hat D_{\lambda} || 0^+_S \rangle| $,
and $\beta$ represents the scale of the coordinate.

The calculation of the matrix elements of these operators
presents a formidable task, since it involves matrix elements
of exponentiated operators. However, since the operator
$\hat D_{\lambda,z}$ is a generator of $U(7)$, the matrix elements of
$\hat U$ are the group elements of $U(7)$ (the generalization of the
Wigner ${\cal D}$-functions of the rotation group) and hence
can be evaluated exactly in a basis provided by the irreducible
representation $[N]$ of $U(7)$. A computer program has been written to do
this evaluation numerically, but in the limit of a large model space
($N \rightarrow \infty$) the results can also be obtained in closed form.
Using harmonic oscillator wave functions one recovers the familiar
expressions of the nonrelativistic harmonic oscillator quark model
\cite{copley,closeli} (see Table~VII of Ref. \cite{bil}).
Explicit analytic results can also be obtained for the collective oblate
top, for which partial results were presented in Tables~VIII and~IX
of Ref. \cite{bil}. Here we present the complete results.

The evaluation of the collective form factors proceeds along the lines of
the appendices of Ref. \cite{bil}.
We first evaluate the matrix elements of the operators in Eq.~(\ref{emop})
between initial and final states which corresponds to the case
in which the charge and magnetization are concentrated at the end
points of the string of Fig.~1.
These matrix elements are expressed in terms of spherical Bessel functions
$j_{L}(k\beta)$, and are given in Table~\ref{otff}.
This table, which completes Table~VIII of Ref. \cite{bil}, forms the
backbone from which form factors for
collective models are built. We note that form factors depend only on two
quantities: the scale of the coordinate $\beta$ and the quantity $R$ that
measures the collectivity (see Appendix~B of Ref. \cite{bil}). The latter
quantity appears only in the transition form factors to the vibrational
excitations of the string. In the nonrelativistic limit the resonances
with $[20,1^+]_{(0,0);0}$ and $[70,2^-]_{(0,0);1}$ are decoupled from
the nucleon ground state $[56,0^+]_{(0,0);0}$.

Table~\ref{otff} can be used to study form factors in collective
models of the nucleon.
A collective model of the nucleon is defined here as an object with the
geometric shape of Fig.~1 and with a specified distribution of charge and
magnetization. We consider, in particular, the model specified by the
(normalized) distribution
\ba
g(\beta) &=& \beta^2 \, \mbox{e}^{-\beta/a} \, /2a^3 ~,  \label{prob}
\ea
where $a$ is a scale parameter. The collective form factors are obtained
by folding the matrix elements of $\hat U$ and
$\hat T_m$ with this probability distribution
\ba
{\cal F}(k)   &=& \int \mbox{d} \beta \, g(\beta) \,
\langle \psi_f | \hat U   | \psi_i \rangle ~,
\nonumber\\
{\cal G}_m(k) &=& \int \mbox{d} \beta \, g(\beta) \,
\langle \psi_f | \hat T_m | \psi_i \rangle  ~.
\label{radint}
\ea
Here $\psi$ denotes the spatial part of the baryon wave function.
According to Table~\ref{otff},
for large $N$ the elastic spatial matrix element of
$\hat U$ is given by the spherical Bessel function $j_{0}(k\beta)$.
The ansatz of Eq.~(\ref{prob}) for the probability distribution
is made to obtain the dipole form for the elastic form factor
\ba
{\cal F}(k)=\int \mbox{d} \beta\, g(\beta) j_0(k\beta)=1/(1+k^2a^2)^2 ~.
\ea
Closed expressions for selected collective transition form factors of the
distributed string are given in Table~\ref{cff}, which completes Table~IX
of Ref. \cite{bil}. It is instructive to study both the small and large
$k$ dependence of the form factors.
This dependence is given in Table~\ref{momtr}. For small values
of $k$ the transition form factors ${\cal F}(k)$ behave as
$\sim k^L$ for rotational excitations with $v=v_1+v_2=0$ and orbital angular
momentum $L$, and as $\sim k^2$ for vibrational excitations with
$v=1$ and $L=0$. More interestingly,
for large values of $k$, all form factors drop as powers of $k$.
This property is well-known experimentally and is in contrast with
harmonic oscillator quark models in which all form factors fall off
exponentially \cite{copley,ki,closeli}.
The elastic form factor ${\cal F}(k)$ drops as $k^{-4}$ (by construction),
whereas the transition form factors for all rotational excitations with
$v=v_1+v_2=0$ drop as $k^{-3}$. For vibrational excitations with
$v=1$ and $L=0$, it drops as $k^{-4}$.
The form factors ${\cal G}_{m}(k)$ drop as the derivatives of
${\cal F}(k)$.

\section{Experimental Observables}
\setcounter{equation}{0}

The form factors of Section 3 can be used to calculate quantities which
can be measured. We begin with the elastic electric and magnetic
form factors of the  nucleon.
The elastic collective form factors are given by
\ba
G^{N}_E &=& 3 \int \mbox{d} \beta \, g(\beta)\,
\langle \, \Psi ;\, M_J=1/2 \,| \,
e_3\, \hat U \, | \, \Psi ;\,
M_J=1/2 \, \rangle ~,
\nonumber\\
G^{N}_M &=& 3 \int \mbox{d} \beta \, g(\beta)\,
\langle \, \Psi ;\, M_J=1/2 \, | \,\mu_3\,
e_3\, \sigma_{3,z}\, \hat U \, | \, \Psi ;\,
M_J=1/2 \, \rangle ~,
\ea
where $\Psi$ denotes the nucleon wave function
$^28^N_{1/2}[56,0^+]_{(0,0);0}$ with $N=p,n$. Further $e_3$,
$\mu_3=eg_3/2m_{3}\,$, $m_3$, $g_3$, $s_3=\sigma_3/2$ are the charge
(in units of $e$: $e_u=2/3$, $e_d=-1/3$), scale magnetic moment, mass,
$g$-factor and spin, respectively, of the third constituent.
Using the results of Table~\ref{cff} we obtain
\ba
G_E^p &=& \frac{1}{(1+k^2a^2)^2} ~,
\nonumber\\
G_E^n &=& 0 ~, \label{gepn}
\ea
for the charge form factors. The corresponding proton charge
radius is found to be
\ba
\langle r^2 \rangle_{E}^{p} &=& 12a^2 ~. \label{pchr}
\ea
Similarly, we obtain for the magnetic form factors
\ba
G_M^p &=& \frac{\mu}{(1+k^2a^2)^2} ~,
\nonumber\\
G_M^n &=& \frac{-2\mu}{3(1+k^2a^2)^2} ~. \label{gmpn}
\ea
The corresponding magnetic moments are
\ba
\mu_p &=& \mu ~,
\nonumber\\
\mu_n &=& -2 \mu/3 ~,
\ea
respectively.
Here we have assumed that the mass and the $g$-factor of
the up ($u$) and down ($d$) constituents are identical, $m_u=m_d=m_q$ and
$g_u=g_d=g$. Hence $\mu_u=\mu_d=\mu$ in Eq.~(\ref{gmpn}) is given by
$\mu = e\,g/2m_q$.
The proton and neutron magnetic radii are identical to
the proton charge radius of Eq.~(\ref{pchr}).
The form factors in Eqs.~(\ref{gepn}) and~(\ref{gmpn})
satisfy $G_M^p = \mu G_E^p$ and obey the $SU_{sf}(6)$
relations $G_E^n = 0$ and $G_M^n/G_M^p = -2/3$.

Other (observable) quantities of interest are the helicity amplitudes
in photo- and electroproduction.
The transverse helicity amplitudes
between the initial (ground) state of the nucleon and the final
(excited) state of a baryon resonance are expressed as \cite{bil}
\ba
A^{N}_{\nu} &=& 6 \sqrt{\frac{\pi}{k_0}} \, \Bigl [ \, k
\langle L,0;S,\nu   | J,\nu \rangle \, {\cal B} -
\langle L,1;S,\nu-1 | J,\nu \rangle \, {\cal A} \, \Bigr ] ~,
\label{helamp}
\ea
where $\nu=1/2$, $3/2$ indicates the helicity.
The orbit- and spin-flip amplitudes (${\cal A}$ and ${\cal B}$,
respectively) are given by
\ba
{\cal B} &=& \int \mbox{d} \beta \, g(\beta)\,
\langle \Psi_f;M_J=\nu | \,\mu_3\,
e_3\, s_{3,+}\, \hat U \, | \Psi_i;M_J^{\prime}=\nu-1 \rangle ~,
\nonumber\\
{\cal A} &=& \int \mbox{d} \beta \, g(\beta)\,
 \langle \Psi_f;M_J=\nu | \, \mu_3\,
e_3\, \hat T_{+}/g_3 \, | \Psi_i;M_J^{\prime}=\nu-1 \rangle ~.
\label{ab}
\ea
Here $| \Psi_i \rangle$ denotes the (space-spin-flavor) wave function
of the initial nucleon with
\hfil\break
$^{2}8^N_{1/2}[56,0^+]_{(0,0);0}$ and $N=p,n$, and, similarly,
$| \Psi_f \rangle$ that of the final baryon resonance.
The helicity amplitudes extracted from experiment include the
sign of the subsequent strong decay into the $\pi N$ channel
and an extra conventional sign -- (+) for nucleon (delta) resonances
\cite{capstick}. Therefore, to compare with the experimental results,
we multiply the helicity amplitudes of
Eq.~(\ref{helamp}) with a coefficient
$\zeta = - \mbox{sign}(N^{\ast}\rightarrow N\pi)$ for nucleon resonances
and $\zeta = + \mbox{sign}(\Delta^{\ast}\rightarrow N\pi)$ for delta
resonances \cite{ki}. Although the extraction of this sign and of the
resultant helicity amplitudes is model dependent, we shall conform in this
article with standard practice and extract the sign from a calculation of
strong decays in a simple model, in which it is assumed that the pion is
emitted from a single constituent and which uses the same collective
wave functions \cite{strong}. The values of $\zeta$, corresponding to the
lowest nucleon and delta resonances, are shown in Tables~\ref{tamp}
and~\ref{tamp2}.

When comparing with the experimental data
one must still choose a reference frame which determines the relation
between the three-momentum $k^2$ and the four-momentum
$Q^2=k^2 - k_{0}^2$.
It is convenient to choose the equal momentum or Breit frame where
\ba
k^2 &=& Q^2 + \frac{(W^2-M^2)^2}{2(M^2+W^2)+Q^2} ~.
\ea
Here $M$ is the nucleon mass, $W$ is the mass of the resonance,
and $-Q^2=k_{0}^2 - k^2$ can be interpreted as the mass squared of
the virtual photon. For elastic scattering we have $k^2 = Q^2$~.

If we assume $m_u=m_d=m_q$ and $g_u=g_d=g$ then, just as in the case of
the nucleon electric and magnetic form factors, $\mu_u=\mu_d=\mu$ in
Eq.~(\ref{ab}). In general, the ${\cal B}$ and ${\cal A}$ amplitudes
of Eq.~(\ref{ab}) are proportional to the collective form factors
${\cal F}$ and ${\cal G}_+$ of Eq.~(\ref{radint}), respectively.
Explicit expressions for the helicity amplitudes of Eq.~(\ref{helamp})
can be obtained by combining the corresponding entries
of Table~\ref{cff} with the appropriate spin-flavor
matrix elements \cite{bil}. Some of these are given
in Tables~\ref{tamp} and~\ref{tamp2}.

\section{Breaking of Spin-Flavor Symmetry}
\setcounter{equation}{0}

In the preceeding sections we have assumed $SU_{sf}(6)$ spin-flavor symmetry.
This leads to $G_E^n = 0$ and $G_M^n/G_M^p=-2/3$ for all values of the
momentum transfer, which is not obeyed by the experimental data.
Within a truncated three-constituents configuration space,
in order to have a nonvanishing neutron electric form factor,
as experimentally observed, one must break $SU_{sf}(6)$ \cite{friar}. 
This breaking can be achieved in various ways, {\it e.g.}
by including in the mass operator a hyperfine interaction \cite{iks}, or
by breaking the $D_3$ spatial symmetry allowing
for a quark-diquark structure \cite{tzeng} and flavor-dependent
mass terms. Within the model discussed here (an effective model with
three constituent parts), we study the breaking of
the $SU_{sf}(6)$ symmetry by assuming a flavor-dependent distribution
of the charge and the magnetization along the strings of Fig.~1,
\ba
g_u(\beta) &=& \beta^2 \, \mbox{e}^{-\beta/a_u} /2a_u^3 ~,
\nonumber\\
g_d(\beta) &=& \beta^2 \, \mbox{e}^{-\beta/a_d} /2a_d^3 ~.
\ea
With this dependence, the electric nucleon form factors become
\ba
G_E^p &=& \frac{2e_u}{(1+k^2a_u^2)^2} + \frac{e_d}{(1+k^2a_d^2)^2} ~,
\nonumber\\
G_E^n &=& \frac{2e_d}{(1+k^2a_d^2)^2} + \frac{e_u}{(1+k^2a_u^2)^2} ~.
\label{elff}
\ea
The corresponding proton and neutron charge radii are given by
\ba
\langle r^2 \rangle_{E}^{p} &=& 12 (2 e_u a_u^2 + e_d a_d^2) ~,
\nonumber\\
\langle r^2 \rangle_{E}^{n} &=& 12 (2 e_d a_d^2 + e_u a_u^2) ~.
\ea
In the limit $k\rightarrow\infty$, the electric form factors behave as
\ba
G_E^p &\rightarrow& \frac{1}{k^4} \left[ \frac{2e_u}{a_u^4} +
\frac{e_d}{a_d^4} \right] ~,
\nonumber\\
G_E^n &\rightarrow& \frac{1}{k^4} \left[ \frac{2e_d}{a_d^4} +
\frac{e_u}{a_u^4} \right] ~.
\label{elfflargek}
\ea

If the length of the string in Fig.~1 is slightly different for $u$ and $d$,
so is their mass and thus in principle, their magnetic moment.
Applying the same procedure to the magnetic form factors gives
\ba
G_M^p &=& \frac{4 \mu_u e_u}{3(1+k^2a_u^2)^2}
- \frac{\mu_d e_d}{3(1+k^2a_d^2)^2} ~,
\nonumber\\
G_M^n &=& \frac{4 \mu_d e_d}{3(1+k^2a_d^2)^2}
- \frac{\mu_u e_u}{3(1+k^2a_u^2)^2} ~.
\label{magff}
\ea
where $\mu_u e_u$ and $\mu_d e_d$ are the magnetic moments of the $u$ and
$d$ constituents. The proton and neutron magnetic moments are now
\ba
\mu_p &=& (4\mu_u e_u - \mu_d e_d)/3 ~,
\nonumber\\
\mu_n &=& (4\mu_d e_d - \mu_u e_u)/3 ~,
\ea
and the proton and neutron magnetic radii are given by
\ba
\langle r^2 \rangle_{M}^{p} &=&
12 (4 \mu_u e_u a_u^2 - \mu_d e_d a_d^2)/(4 \mu_u e_u - \mu_d e_d) ~,
\nonumber\\
\langle r^2 \rangle_{M}^{n} &=&
12 (4 \mu_d e_d a_d^2 - \mu_u e_u a_u^2)/(4 \mu_d e_d - \mu_u e_u) ~.
\label{radmag}
\ea
The asymptotic limit ($k^2\rightarrow\infty$) of the magnetic form
factors is
\ba
G_M^p &\rightarrow& {1\over k^4}\Bigl [ {4\mu_u e_u\over 3a_u^4}
- {\mu_d e_d\over 3a_d^4} \Bigl ] ~,
\nonumber\\
G_M^n &\rightarrow& {1\over k^4}\Bigl [ {4\mu_d e_d\over 3a_d^4}
- {\mu_u e_u\over 3a_u^4} \Bigl ] ~.
\label{magfflargek}
\ea

We note at this stage that if the masses of the up and down constituents are
slightly different, $S_3$ ($D_3$) symmetry is also broken in the wave functions
and spectrum, causing a splitting of the degenerate rotations and vibrations.
This effect will be analyzed in detail when studying strange baryons where it
is much larger due to the large difference in the mass of the strange
constituent relative to that of the up and down constituents.
(We also note that our main interest is to present results for observable
quantities due to spin-flavor breaking in a truncated space, 
independently from its magnitude.
Different QCD spin flavor mechanisms give different values for the effective
masses, $m_d$ and $m_u$, magnetic moments $\mu_d$ and $\mu_u$, and sizes $a_u$
and $a_d$, both with $a_u < a_d$ and $a_u > a_d$ \cite{cmt}.)

The breaking of spin-flavor symmetry has also influence on the helicity
amplitudes. Inserting the appropriate spin-flavor coefficients in
Eq.~(\ref{ab}) one obtains the results for the orbit- and spin-flip
amplitudes, ${\cal A}$ and ${\cal B}$, given in Tables~\ref{nucleon}
and~\ref{delta}. The helicity amplitudes of Eq.~(\ref{helamp})
are now given in terms of the flavor-dependent collective form factors
${\cal F}_u(k)$, ${\cal G}_{u,+}(k)$ and ${\cal F}_d(k)$,
${\cal G}_{d,+}(k)$, which depend on the size parameters,
$a_u$ and $a_d$, respectively. Explicit expressions for the various
helicity amplitudes are available on request.

Table~\ref{nucleon} shows that two sets of helicity amplitudes which
were previously zero due to spin-flavor symmetry, are nonvanishing
in the presence of flavor-dependent distributions: (i) the Moorehouse
selection rule for the proton helicity amplitudes for the
$^{4}8_{J}[70,L]$ resonances is broken, and (ii) the neutron helicity-3/2
amplitudes for the $^{2}8_{J}[56,L]$ resonances are nonvanishing.

\section{Stretchable Strings}
\setcounter{equation}{0}

In a string-like model of hadrons one expects on the basis of QCD
\cite{johnson,bars} that strings will elongate (hadrons swell)
as their energy increases. This effect can be
easily included in the present analysis by making the scale parameters
of the strings energy dependent. In order to study the swelling of hadrons
with increasing excitation energy, we use here the simple ansatz
\ba
a &=& a_0\Bigl ( 1 + \xi\,{W-M\over M}\Bigr ) ~, \label{stretch}
\ea
where $M$ is the nucleon mass and $W$ the resonance mass. This ansatz
introduces a new parameter, the stretchability of the string, $\xi$. The
arguments of Ref. \cite{johnson} and the analysis of the experimental mass
spectrum (Regge trajectories) suggest $\xi\approx 1$.
Spin-flavor $SU_{sf}(6)$ symmetry breaking may also effect the value
of $\xi$, but this is likely to be a higher order effect. Hence we
parameterize the breaking as
\ba
a_u &=& a_{u,0}\Bigl ( 1 + \xi\,{W-M\over M}\Bigr ) ~,
\nonumber\\
a_d &=& a_{d,0}\Bigl ( 1 + \xi\,{W-M\over M}\Bigr ) ~.
\ea
{\it i.e.} we assume the stretchability to be flavor independent.

\section{Analysis of Experimental Data}
\setcounter{equation}{0}
\subsection{Spin-Flavor Breaking}

In this section we investigate the effect of the flavor dependence
on the elastic and transition form factors of nonstrange baryons.
We begin by discussing the determination of the parameters. For all cases
we take $g_u=g_d=1$.
For the calculations in which the $SU_{sf}(6)$ symmetry
is satisfied ($\mu_u=\mu_d=\mu$ and $a_u=a_d=a$), we determine the
scale magnetic moment $\mu$ from the proton magnetic moment
$\mu=\mu_p= 2.793$ $\mu_N$, which corresponds to a constituent mass
of $m_u=m_d=0.336$ GeV.
Since the values of the helicity amplitudes $A^{N}_{\nu}$ are usually
given in GeV$^{-1/2}$, we express the scale magnetic moment appearing
in Eq.~(\ref{ab}) in units of $\mu = 0.127$ GeV$^{-1}$ ($\hbar=c=1$).
In \cite{bil} the scale parameter $a$ was determined from the proton
charge radius (see Eq.~(\ref{pchr})). Here we prefer to use a
simultaneous fit to the proton and neutron charge radii
\cite{simon,krohn}, to the proton electric and magnetic form factors up to
$Q^2=5$ (GeV/c)$^{2}$ \cite{walker}, and to the neutron electric
\cite{platchkov} and magnetic form factors \cite{lung,bruins,bartel}
up to $Q^2=4$ (GeV/c)${^2}$. As a result we find $a=0.232$ fm.

In order to study the sensitivity of the form factors (elastic and transition)
to breaking of $SU_{sf}(6)$ symmetry, we assume that the constituent masses
$m_u$ and $m_d$ are determined from the magnetic moments with quark 
$g$-factors $g_u=g_d=1$. Using $\mu_u=2.777\mu_{N}$ and $\mu_d=2.915\mu_{N}$,
we find $m_u=0.338$ GeV and $m_d=0.322$ GeV, respectively. The scale
parameters $a_u$ and $a_d$ are determined from a simultaneous fit to the
proton and neutron charge radii and the proton and neutron electric and
magnetic form factors, $a_u=0.230$ fm and $a_d=0.257$ fm.
We note that the magnitude of the breaking
both in the effective masses and scales so determined is too large when
compared with estimates based on the $m_u-m_d$ mass difference of the
``current'' quarks and on QCD perturbation estimates with $\alpha_{s}=0.5$
\cite{cmt}. The necessity to use $a_u\neq a_d$ in the present model should
be interpreted as a consequence of the truncation of configuration space
to the pure three-constituents states. 
Our purpose, however, is to understand what happens to the form
factors when one breaks $SU_{sf}(6)$ in the truncated space.
Since the $a_u$ and $a_d$ are effective quantities that incorporate all
complexities of the non-three-constituents configurations, they may have
significant final-state dependence (which is ignored in the present study).

We first discuss the elastic form factors. Figs.~2 and~3 show the electric
form factors of the neutron and the proton divided by the dipole form,
$F_D = 1/(1+Q^2/0.71)^2$. The division by $F_D$ emphasizes the
effect of the breaking of spin-flavor symmetry. Figs.~4
and~5 show the results for the neutron and proton magnetic form factors,
respectively. We see that while the breaking of spin-flavor symmetry can
account for the non-zero value of $G_E^n$ and gives a good description
of the data, it worsens the fit to the proton electric and neutron
magnetic form factors.
This implies that, in addition to not being of the right order of magnitude
when compared with QCD estimates, the simple mechanism for spin-flavor
breaking discussed in Sect. 5 does not produce the right phenomenology and
other contributions, such as polarization of the neutron into
$p+\pi^{-}$, play an important role in the neutron electric form factor
\cite{cmt}.
(A coupling to the meson
cloud through $\rho$, $\omega$ and $\phi$ mesons is indeed expected to
contribute in this range of $Q^2$, see Fig.~1 of Ref. \cite{vmd}.)
This conclusion (i.e. worsening the proton form factors)
applies also to the other mechanisms of spin-flavor symmetry
breaking mentioned above, such as that induced by a hyperfine interaction
\cite{iks} which gives $a_u < a_d$ (`moves the up quark to the center and
the down quark to the periphery') although it was not discussed in 
Ref. \cite{iks}. This pattern is a
consequence of the fact that within the framework of constituent models
$G_E^p$, $G_E^n$, $G^p_M$ and $G_M^n$ are intertwined.

We note in passing that spin-flavor breaking also alters the ratio
of the magnetic form factors $G_M^n/G_M^p$. From Eq.~(\ref{magfflargek})
we find that for $k^2 \rightarrow \infty$ this ratio approaches
$G_M^n/G_M^p \rightarrow (4 \mu_d e_d a_u^4 - \mu_u e_u a_d^4)/
(4 \mu_u e_u a_d^4 - \mu_d e_d a_u^4)$.
With the values of $a_u$, $a_d$ and of $\mu_u$, $\mu_d$ given above,
we calculate this ratio to be $-0.541$. On the basis of perturbative QCD
the ratio is expected to approach
$-1/2+{\cal O}(\mbox{ln }Q^2)$ for large values of $Q^2$ \cite{warns}.
With harmonic oscillator form factors this ratio approaches $-1/4$.
Without the breaking of the spin-flavor symmetry this ratio is
$-2/3$ independent of $Q^2$ for both the collective and the harmonic
oscillator case. The breaking of spin-flavor symmetry brings the value
of the ratio for $Q^2\rightarrow\infty$ closer to the p-QCD value. From
Fig.~6 we can see that the experimental situation does not
show any indication that the perturbative regime has been reached,
at least up to $Q^2 \leq 3$ (GeV/c)$^2$.

Next we discuss the transverse helicity amplitudes $A_{1/2}$
and $A_{3/2}$. The results of the calculations with and without
spin-flavor breaking are shown in Figs.~7-11 for nucleon resonances
and in Figs.~12-15 for delta resonances. From
these figures it is seen that the effect is rather small.
Only in those cases in which the amplitude with $SU_{sf}(6)$
symmetry is zero, the effect is of some relevance.
Such is the case for the neutron amplitude $A_{3/2}^{n}$ of the
$N(1680)F_{15}$ resonance shown in Fig.~11, and for the
proton amplitudes of the $N(1650)S_{11}$ (see Fig.~9),
$N(1675)D_{15}$ and $N(1700)D_{13}$ resonances, which all belong to
the $^{4}8_{J}[70,L=1^{-}]$ multiplet.
The small effect of the spin-flavor symmetry breaking is emphasized
in Figs.~16-18 where the helicity asymmetries
\ba
A &=& {A_{1/2}^2 - A_{3/2}^2\over A_{1/2}^2 + A_{3/2}^2}
\ea
are plotted versus $Q^2$. The conclusion that one can draw
from this analysis is that, for all purposes, except the electric
form factor of the neutron, the breaking of spin-flavor symmetry
according to the mechanism of Sect.~5 is of little importance.
As an additional comment, we note that in Figs.~8 and~9 we have shown
only the amplitudes with no mixing, $\theta=0^{\circ}$
(see Eq. (10.3) of Ref. \cite{bil}),
since our purpose is that of displaying the effects
of spin-flavor breaking induced by $a_u < a_d$.
(The mixing between the two $S_{11}$ states may be effected by meson
cloud corrections, specifically, $N$-$\eta$ contributions.)

The helicity amplitudes shown in Figs.~7-15 all describe rotational
excitations in the collective model.
It is of interest to comment briefly on vibrational
excitations. As one can see from Tables~\ref{cff} and~\ref{tamp},
the matrix elements of transition operators to the states
$[56,0^+]_{(1,0);0}$ and $[70,0^+]_{(0,1);0}$ vanish in the large $N$
limit of the collective model (and $R^2 \neq 0$).
Nevertheless, it is instructive to study these matrix elements for
finite $N$ (but large) and $R^2 \neq 0,1$.
Denoting by $\chi = (1-R^2)/R\sqrt{N}$ the strength of the coupling, we
show in Fig.~19 the corresponding transverse helicity amplitude for
$N(1440)P_{11}$ (the Roper resonance). We note that the calculated
amplitude has the opposite sign of the experimental amplitude
(just as in \cite{capstick}, as well as in the harmonic oscillator limit
of the algebraic model \cite{bil}).
However, the behavior of the amplitude with $Q^2$ is particular
enough to be able to say something concerning the nature of the
Roper resonance once more accurate data will be available.

\subsection{Stretching}
\setcounter{equation}{0}

In this section we analyze what happens to the helicity amplitudes
with the stretching mechanism of Sect.~6. Figs.~20-22 show the effect
of stretching on the helicity amplitudes for $\Delta(1232)P_{33}$,
$N(1520)D_{13}$ and $N(1680)F_{15}$. It is seen that the effect of
stretching, especially if one takes the value $\xi\approx 1$ suggested
by the arguments of \cite{johnson} and the Regge behavior of nucleon
resonances (see {\it e.g.} Fig.~5 of \cite{bil}), is rather large.
In particular, the data for $N(1520)D_{13}$ and $N(1680)F_{15}$
show a clear indication that the form
factors are dropping faster than expected on the basis of the dipole
form. (Of course for the elastic form actors there is no stretching.)
We suggest that future data at CEBAF and MAMI be used to analyze
the effects of stretching on the helicity amplitudes.

\section{Summary and conclusions}
\setcounter{equation}{0}

In this article, we have exploited the algebraic approach to baryon
structure introduced in \cite{bil} to analyze simultaneously elastic
form factors and helicity amplitudes in photo- and electroproduction.
The use of algebraic methods allows us to study different situations,
such as the harmonic oscillator quark model and the collective model,
within the same framework. The logic of the
method is that, by starting from the charge and magnetization distribution
of the ground state (assuming a dipole form to the elastic form factor of
the nucleon), one can obtain the transition form factors to the excited
states. In the `collective' model, this procedure
yields {\it a power dependence of all form factors} (elastic
and inelastic) on $Q^2$.
We have analyzed two aspects of hadronic structure: (i) the
breaking of $SU_{sf}(6)$ symmetry, 
and (ii) the stretching of hadrons
with increasing excitation energy. We find that, whereas the breaking
of the spin-flavor symmetry hardly effects the helicity amplitudes,
the stretching of hadrons does have a noticeable influence.

The disagreement between experimental and theoretical elastic form
factors and helicity amplitudes in the low-$Q^2$ region
$0\leq Q^2\leq 1$ (GeV/c)$^2$, requires further investigation.
We think that this disagreement is due to coupling of the photon
to the meson cloud, ({\it i.e.} configurations of the type
$q^3-q\bar{q}$). In the case of the elastic form factors, these
effects were in part analyzed in vector dominance models \cite{vmd} by
writing the amplitude as the sum of two terms. We think that this
analysis (which was done before the advent of quark model calculations)
should be repeated by using for the `intrinsic' part the constituent
form factors discussed in this article. Coupling of the photon to
$\rho$ (isovector), $\omega$ and $\phi$ (isoscalar) vector mesons
can produce a non-zero neutron form factor which describes the data
without worsening the proton form factor description. For the helicity
amplitudes, the effects could either be calculated directly \cite{lee92},
or be parametrized by meson (not necessarily vector) dominance models.
We note however that in either case, since configurations of the type
$q^3-q\bar{q}$ have much larger spatial extent than $q^3$, these effects
are expected to drop faster with momentum
transfer $Q^2$ than the constituent form factors.
Also, since meson exchange corrections contribute differently to different
channels, this effect will be state dependent.

Another aspect that requires further investigation is the contribution of
the spin-orbit and non-additive part in the transition operators. Since the
algebraic formulation is now in place, these effects can be investigated.
The corresponding results will be reported elsewhere.

\section*{Acknowledgements}

This work is supported in part
by CONACyT, M\'exico under project 400340-5-3401E, DGAPA-UNAM under
project IN105194 (R.B.), by D.O.E. Grant DE-FG02-91ER40608 (F.I.) and
by grant No. 94-00059 from the United States-Israel Binational Science
Foundation (BSF), Jerusalem, Israel (A.L.).

\clearpage
\begin{table}
\centering
\caption[Oblate top form factors]{\small
Matrix elements of the transition operators
of Eq.~(\ref{emop}) in the large $N$ limit.
The final states are labeled by $[dim,L^P]_{(v_1,v_2);K}$,
where $dim$
denotes the dimension of the $SU_{sf}(6)$ representation.
The initial state is $[56,0^{+}]_{(0,0);0}$.
\normalsize} \label{otff}
\vspace{15pt}
\begin{tabular}{cccc}
\hline
& & & \\
Final state & $\langle \psi_f | \hat U | \psi_i \rangle$
& $\langle \psi_f | \hat T_{0}   | \psi_i \rangle/m_3 k_0 \beta$
& $\langle \psi_f | \hat T_{\pm} | \psi_i \rangle/m_3 k_0 \beta$ \\
& & & \\
\hline
& & & \\
$[56,0^+]_{(0,0);0}$ & $j_0(k \beta)$ & $j_1(k \beta)$ & 0 \\
& & & \\
$[20,1^+]_{(0,0);0}$ & 0 & 0 & 0 \\
& & & \\
$[70,1^-]_{(0,0);1}$ & $-i \, \sqrt{3} \, j_1(k \beta)$
& $i \frac{1}{\sqrt{3}} [j_0(k \beta)-2j_2(k \beta)]$
& $\mp i \, \sqrt{\frac{2}{3}} [j_0(k \beta)+ j_2(k \beta)]$ \\
& & & \\
$[56,2^+]_{(0,0);0}$ & $\frac{1}{2} \sqrt{5} \, j_2(k \beta)$
& $-\frac{1}{2\sqrt{5}} [2j_1(k \beta)-3j_3(k \beta)]$
& $\pm \sqrt{\frac{3}{10}} [j_1(k \beta)+j_3(k \beta)]$ \\
& & & \\
$[70,2^-]_{(0,0);1}$ & 0 & 0 & 0 \\
& & & \\
$[70,2^+]_{(0,0);2}$ & $-\frac{1}{2} \sqrt{15} \, j_2(k \beta)$
& $\frac{1}{2} \sqrt{\frac{3}{5}} \, [2j_1(k \beta)-3j_3(k \beta)]$
& $\mp 3 \sqrt{\frac{1}{10}} \, [j_1(k \beta)+j_3(k \beta)]$ \\
& & & \\
$[56,0^+]_{(1,0);0}$
& $-\frac{1-R^2}{2R \sqrt{N}} \, k \beta \, j_1(k \beta)$
& $\frac{1-R^2}{6R \sqrt{N}} \, k \beta [2j_0(k \beta)-j_2(k \beta)]$
& 0 \\
& & & \\
$[70,0^+]_{(0,1);0}$
& $\frac{\sqrt{1+R^2}}{2R \sqrt{N}} \, k \beta \, j_1(k \beta)$
& $-\frac{\sqrt{1+R^2}}{6R \sqrt{N}} \,
k \beta [2j_0(k \beta)-j_2(k \beta)]$ & 0 \\
& & & \\
\hline
\end{tabular}
\end{table}

\clearpage
\begin{table}
\centering
\caption[Collective form factors]{\small
Collective form factors in the large $N$ limit.
$H(x)=\arctan x - x/(1+x^2)$. Notation as in Table~\ref{otff}.
\normalsize} \label{cff}
\vspace{15pt}
\begin{tabular}{cccc}
\hline
& & & \\
Final state & ${\cal F}(k)$
& ${\cal G}_{0}(k)/m_3 k_0 a$
& ${\cal G}_{\pm}(k)/m_3 k_0 a$ \\
& & & \\
\hline
& & & \\
$[56,0^+]_{(0,0);0}$
& $\frac{1}{(1+k^2a^2)^2}$ & $\frac{4ka}{(1+k^2a^2)^3}$ & 0  \\
& & & \\
$[20,1^+]_{(0,0);0}$ & 0 & 0 & 0 \\
& & & \\
$[70,1^-]_{(0,0);1}$ & $-i \, \sqrt{3} \, \frac{ka}{(1+k^2a^2)^2}$
& $i \, \sqrt{3} \, \frac{1-3k^2a^2}{(1+k^2a^2)^3}$
& $\mp i \, \sqrt{6} \, \frac{1}{(1+k^2a^2)^2}$ \\
& & & \\
$[56,2^+]_{(0,0);0}$
& $ \frac{1}{2} \sqrt{5}\left[ \frac{-1}{(1+k^2a^2)^2} \right.$
& $-\frac{1}{2} \sqrt{5}\left[ \frac{3+7k^2a^2}{ka(1+k^2a^2)^3} \right.$
& $\pm \sqrt{\frac{15}{2}}\left[ \frac{-1}{ka(1+k^2a^2)^2} \right.$ \\
& $\left. \hspace{1cm} + \frac{3}{2k^3a^3} H(ka) \right]$
& $\left. \hspace{1cm} - \frac{9}{2k^4a^4} H(ka) \right]$
& $\left. \hspace{1cm} + \frac{3}{2k^4a^4} H(ka) \right]$ \\
& & & \\
$[70,2^-]_{(0,0);1}$ & 0 & 0 & 0 \\
& & & \\
$[70,2^+]_{(0,0);2}$
& $-\frac{1}{2} \sqrt{15}\left[ \frac{-1}{(1+k^2a^2)^2} \right.$
& $ \frac{1}{2} \sqrt{15}\left[ \frac{3+7k^2a^2}{ka(1+k^2a^2)^3} \right.$
& $\mp \frac{3}{2} \sqrt{10}\left[ \frac{-1}{ka(1+k^2a^2)^2} \right.$ \\
& $\left. \hspace{1cm} + \frac{3}{2k^3a^3} H(ka) \right]$
& $\left. \hspace{1cm} - \frac{9}{2k^4a^4} H(ka) \right]$
& $\left. \hspace{1cm} + \frac{3}{2k^4a^4} H(ka) \right]$ \\
& & & \\
$[56,0^+]_{(1,0);0}$
& $- \frac{1-R^2}{R \sqrt{N}} \frac{2k^2a^2}{(1+k^2a^2)^3}$
& $  \frac{1-R^2}{R \sqrt{N}} \frac{4ka(1-2k^2a^2)}{(1+k^2a^2)^4}$ & 0 \\
& & & \\
$[70,0^+]_{(0,1);0}$
& $ \frac{\sqrt{1+R^2}}{R\sqrt{N}} \frac{2k^2a^2}{(1+k^2a^2)^3}$
& $-\frac{\sqrt{1+R^2}}{R\sqrt{N}} \frac{4ka(1-2k^2a^2)}{(1+k^2a^2)^4}$
& 0 \\
& & & \\
\hline
\end{tabular}
\end{table}

\clearpage
\begin{table}
\centering
\caption[Small and large momentum transfer]{\small
Behavior of the collective form factors of Table~\ref{cff} for
$ka \ll 1$ and $ka \gg 1$.
\normalsize} \label{momtr}
\vspace{15pt}
\begin{tabular}{c|ccc|ccc}
\hline
& & & & & & \\
& \multicolumn{3}{c|} {$ka \ll 1$} & \multicolumn{3}{c} {$ka \gg 1$} \\
& & & & & & \\
Final state & ${\cal F}(k)$ & ${\cal G}_{0}(k)/m_3 k_0 a$
& ${\cal G}_{\pm}(k)/m_3 k_0 a$ & ${\cal F}(k)$
& ${\cal G}_{0}(k)/m_3 k_0 a$ & ${\cal G}_{\pm}(k)/m_3 k_0 a$ \\
& & & & & & \\
\hline
& & & & & & \\
$[56,0^{+}]_{(0,0);0}$ & $\sim 1$ & $\sim ka$ & 0
& $\sim (ka)^{-4}$ & $\sim (ka)^{-5}$ & 0 \\
& & & & & & \\
$[70,1^{-}]_{(0,0);1}$ & $\sim ka$ & $\sim 1$ & $\sim 1$
& $\sim (ka)^{-3}$ & $\sim (ka)^{-4}$ & $\sim (ka)^{-4}$ \\
& & & & & & \\
$[56,2^{+}]_{(0,0);0}$ & $\sim (ka)^2$ & $\sim ka$ & $\sim ka$
& $\sim (ka)^{-3}$ & $\sim (ka)^{-4}$ & $\sim (ka)^{-4}$ \\
& & & & & & \\
$[70,2^{+}]_{(0,0);2}$ & $\sim (ka)^2$ & $\sim ka$ & $\sim ka$
& $\sim (ka)^{-3}$ & $\sim (ka)^{-4}$ & $\sim (ka)^{-4}$ \\
& & & & & & \\
$[56,0^{+}]_{(1,0);0}$ & $\sim (ka)^2$ & $\sim ka$ & 0
& $\sim (ka)^{-4}$ & $\sim (ka)^{-5}$ & 0 \\
& & & & & & \\
$[70,0^{+}]_{(0,1);0}$ & $\sim (ka)^2$ & $\sim ka$ & 0
& $\sim (ka)^{-4}$ & $\sim (ka)^{-5}$ & 0 \\
& & & & & & \\
\hline
\end{tabular}
\end{table}

\clearpage
\begin{table}
\centering
\caption[Transverse helicity amplitudes]{\small
Analytic expressions for the transverse proton helicity
amplitudes of some nucleon resonances, derived using
Eq.~(\ref{helamp}) and Tables~\ref{cff} and~\ref{nucleon}
with $a_u=a_d=a$, $m_u=m_d=m_q$, $g_u=g_d=g$ and $\mu_u=\mu_d=\mu$.
$Z(x) = -\frac{1}{(1+x^2)^2} + \frac{3}{2x^3}H(x)$ with
$H(x) = \arctan x - {x\over (1+x^2)}$ and $\chi=(1-R^2)/R\sqrt{N}$~.
$\zeta$ denotes an additional multiplicative sign-factor in accord
with the convention explained in the text.
\normalsize} \label{tamp}
\vspace{15pt}
\begin{tabular}{ccccc}
\hline
& & & &  \\
Resonance & State & $\nu$ & $A^{p}_{\nu}$ & $\zeta$ \\
& & & & \\
\hline
& & & & \\
$N(1535)S_{11}$ & $^{2}8_{1/2}[70,1^-]_{(0,0);1}$ & $1/2$
& $i\sqrt{2}\sqrt{\pi\over k_0}\mu\frac{1}{(1+k^2a^2)^2}
\Bigl [2{m_qk_0a\over g} + k^2a\Bigr ]$
& $+1$ \\
& & & & \\
$N(1520)D_{13}$ & $^{2}8_{3/2}[70,1^-]_{(0,0);1}$ & $1/2$
& $2i\sqrt{\pi\over k_0}\mu\frac{1}{(1+k^2a^2)^2}
\Bigl [{m_qk_0a\over g} - k^2a\Bigr ]$
& $+1$ \\
%& & & & \\
& & $3/2$ & $2i\sqrt{3}\sqrt{\pi\over k_0}\mu
\frac{1}{(1+k^2a^2)^2}{m_qk_0a\over g}$
& $+1$ \\
& & & & \\
$N(1650)S_{11}$ & $^{4}8_{1/2}[70,1^-]_{(0,0);1}$ & $1/2$ & 0
& \\
& & & & \\
$N(1700)D_{13}$ & $^{4}8_{3/2}[70,1^-]_{(0,0);1}$ & $1/2,\,3/2$ & 0
& \\
& & & & \\
$N(1675)D_{15}$ & $^{4}8_{5/2}[70,1^-]_{(0,0);1}$ & $1/2,\,3/2$ & 0
& \\
& & & & \\
$N(1720)P_{13}$ & $^{2}8_{3/2}[56,2^+]_{(0,0);0}$ & $1/2$
& $-\sqrt{2}\sqrt{\pi\over k_0}\mu
\Bigl [ 3{m_qk_0a\over g} + k^2a\Bigr] {1\over ka} Z(ka)$
& $-1$ \\
%& & & & \\
& & $3/2$ & $\sqrt{6} \sqrt{\pi\over k_0}\mu \,
{m_qk_0a\over g} \, {1\over ka} Z(ka)$
& $-1$ \\
& & & & \\
$N(1680)F_{15}$ & $^{2}8_{5/2}[56,2^+]_{(0,0);0}$ & $1/2$
& $-\sqrt{3}\sqrt{\pi\over k_0}\mu
\Bigl [ 2{m_qk_0a\over g} - k^2a\Bigr] {1\over ka} Z(ka)$
& $-1$ \\
%& & & & \\
& & $3/2$ & $-2\sqrt{6} \sqrt{\pi\over k_0}\mu \,
{m_qk_0a\over g} \, {1\over ka} Z(ka)$
& $-1$ \\
& & & & \\
$N(1440)P_{11}$ & $^{2}8_{1/2}[56,0^+]_{(1,0);0}$ & $1/2$
& $-2 \chi \sqrt{\pi\over k_0}\mu k \, \frac{2k^2a^2}{(1+k^2a^2)^3}$
& $-1$ \\
& & & & \\
\hline
\end{tabular}
\end{table}

\clearpage
\begin{table}
\centering
\caption[Transverse helicity amplitudes]{\small
Analytic expressions for the transverse helicity
amplitudes of some delta resonances resonances, derived using
Eq.~(\ref{helamp}) and Tables~\ref{cff} and~\ref{delta}
with $a_u=a_d=a$, $m_u=m_d=m_q$, $g_u=g_d=g$ and $\mu_u=\mu_d=\mu$.
Notation as in Table~\ref{tamp}.
\normalsize} \label{tamp2}
\vspace{15pt}
\begin{tabular}{ccccc}
\hline
& & & &  \\
Resonance & State & $\nu$ & $A^{p}_{\nu}=A^{n}_{\nu}$ & $\zeta$ \\
& & & & \\
\hline
& & & & \\
$\Delta(1232)P_{33}$ & $^{4}10_{3/2}[56,0^+]_{(0,0);0}$ & $1/2$
& $-{2\sqrt{2}\over 3}\sqrt{\pi\over k_0}\mu k\frac{1}{(1+k^2a^2)^2}$
& $+1$ \\
& & & & \\
& & $3/2$ & $-{2\sqrt{2}\over \sqrt{3}}\sqrt{\pi\over k_0}\mu k
\frac{1}{(1+k^2a^2)^2}$
& $+1$ \\
& & & & \\
$\Delta(1620)S_{31}$ & $^{2}10_{1/2}[70,1^-]_{(0,0);1}$ & $1/2$
& $-i{\sqrt{2}\over 3}\sqrt{\pi\over k_0}\mu\frac{1}{(1+k^2a^2)^2}
\Bigl [6{m_qk_0a\over g} - k^2a\Bigr ]$
& $-1$ \\
& & & & \\
$\Delta(1700)D_{33}$ & $^{2}10_{3/2}[70,1^-]_{(0,0);1}$ & $1/2$
& $-i{2\over 3}\sqrt{\pi\over k_0}\mu\frac{1}{(1+k^2a^2)^2}
\Bigl [3{m_qk_0a\over g} + k^2a\Bigr ]$
& $-1$ \\
& & & & \\
& & $3/2$ & $-2i\sqrt{3}\sqrt{\pi\over k_0}\mu
\frac{1}{(1+k^2a^2)^2}{m_qk_0a\over g}$
& $-1$ \\
& & & & \\
$\Delta(1600)P_{33}$ & $^{4}10_{3/2}[56,0^+]_{(1,0);0}$ & $1/2$
& ${2\sqrt{2}\over 3} \chi \sqrt{\pi\over k_0}\mu k \,
\frac{2k^2a^2}{(1+k^2a^2)^3}$
& $+1$ \\
& & & & \\
& & $3/2$
& ${2\sqrt{2}\over \sqrt{3}} \chi \sqrt{\pi\over k_0}\mu k \,
\frac{2k^2a^2}{(1+k^2a^2)^3}$
& $+1$ \\
& & & & \\
\hline
\end{tabular}
\end{table}

\clearpage
\begin{table}
\centering
\caption[Spin- and orbit-flip amplitudes for nucleon resonances]
{\small Orbit- and spin-flip amplitudes of Eq.~(\ref{ab}), associated
with transverse helicity amplitudes for nucleon
resonances (proton-target couplings) according to Eq.~(\ref{helamp}).
$y_i= x_i/g_i$ and $x_i= \mu_ie_i$;
${\cal F}_i(k)$, ${\cal G}_{+,i}$ are obtained from the corresponding
entries in Table~\ref{cff} with $a \rightarrow a_i$, $m_{3} \rightarrow m_i$
with $i=u,d$. Neutron-target couplings are obtained by interchanging
$u \leftrightarrow d$.
\normalsize}
\label{nucleon}
\vspace{15pt}
\begin{tabular}{cc|cc}
\hline
& & & \\
& & \multicolumn{2}{|c} {$A^p_{\nu}$} \\
State & $\nu$ & ${\cal A}$ & ${\cal B}$ \\
& & & \\
\hline
& & & \\
$^{2}8[56]$ & $1/2$
& $\frac{1}{3} \left[ 2y_u {\cal G}_{u,+}(k) + y_d {\cal G}_{d,+}(k) \right]$
& $\frac{1}{9} \left[ 4x_u {\cal F}_{u}(k) - x_d {\cal F}_{d}(k) \right]$ \\
& $3/2$ & $\frac{1}{3}
\left[ 2y_u {\cal G}_{u,+}(k) + y_d {\cal G}_{d,+}(k) \right]$ & $0$ \\
& & & \\
$^{2}8[70]$ & $1/2$ & $\frac{1}{3\sqrt{2}}
\left[ y_u {\cal G}_{u,+}(k) - y_d {\cal G}_{d,+}(k) \right]$
& $\frac{1}{9\sqrt{2}}
\left[ 5x_u {\cal F}_{u}(k) + x_d {\cal F}_{d}(k) \right]$ \\
& $3/2$ & $\frac{1}{3\sqrt{2}}
\left[ y_u {\cal G}_{u,+}(k) - y_d {\cal G}_{d,+}(k) \right]$ & $0$ \\
& & & \\
$^{2}8[20]$ & $1/2$ & 0 & 0 \\ & $3/2$ & 0 & 0 \\
& & & \\
$^{4}8[70]$ & $1/2$ & 0
& $\frac{1}{9\sqrt{2}}
\left[ x_u {\cal F}_{u}(k) + 2x_d {\cal F}_{d}(k) \right]$ \\
& $3/2$ & 0 & $\frac{1}{3\sqrt{6}}
\left[ x_u {\cal F}_{d}(k) + 2x_d {\cal F}_{d}(k) \right]$ \\
& & & \\
\hline
\end{tabular}
\end{table}

\clearpage
\begin{table}
\centering
\caption[Spin- and orbit-flip amplitudes for delta resonances]{\small
Orbit- and spin-flip amplitudes of Eq.~(\ref{ab}),
associated with transverse helicity
amplitudes for delta resonances. Notation as in Table~\ref{nucleon}.
\normalsize} \label{delta}
\vspace{15pt}
\begin{tabular}{cc|cc}
\hline
& & & \\
& & \multicolumn{2}{|c} {$A^{p}_{\nu}=A^{n}_{\nu}$} \\
State & $\nu$ & ${\cal A}$ & ${\cal B}$ \\
& & & \\
\hline
& & & \\
$^{2}10[70]$ & $1/2$ & $\frac{-1}{3\sqrt{2}}
\left[ y_u {\cal G}_{u,+}(k) - y_d {\cal G}_{d,+}(k) \right]$
& $\frac{ 1}{9\sqrt{2}}
\left[ x_u {\cal F}_{u}(k) - x_d {\cal F}_{d}(k) \right]$ \\
& $3/2$ & $\frac{-1}{3\sqrt{2}}
\left[ y_u {\cal G}_{u,+}(k) - y_d {\cal G}_{d,+}(k) \right]$ & $0$ \\
& & & \\
$^{4}10[56]$ & $1/2$ & 0 & $\frac{-\sqrt{2}}{9}
\left[ x_u {\cal F}_{u}(k) - x_d {\cal F}_{d}(k) \right]$ \\
& $3/2$ & 0 & $\frac{-\sqrt{2}}{3\sqrt{3}}
\left[ x_u {\cal F}_{u}(k) - x_d {\cal F}_{d}(k) \right]$ \\
& & & \\
\hline
\end{tabular}
\end{table}

\clearpage
\section*{Figure Captions}
\noindent
{\bf Figure 1:}
Collective model of baryons and its idealized string configuration
(the charge distribution of the proton is shown as an example).

\vspace{15pt}
\noindent
{\bf Figure 2:}
Comparison between the experimental neutron electric form factor
$G_E^n$, and the collective form factor
with and without flavor breaking (dashed and solid lines).
The experimental data, taken from \cite{platchkov}, and the calculations
are divided by the dipole form factor, $F_D=1/(1+Q^2/0.71)^2$.

\vspace{15pt}
\noindent
{\bf Figure 3:}
Comparison between the experimental proton electric form factor
$G_E^p$, and the collective form factor
with and without flavor breaking (dashed and solid lines).
The experimental data taken from \cite{walker},
and the calculations are divided by the dipole form
factor, $F_D=1/(1+Q^2/0.71)^2$.

\vspace{15pt}
\noindent
{\bf Figure 4:}
Comparison between the experimental neutron magnetic form factor
$G_M^n$, and the collective form factor
with and without flavor breaking (dashed and solid lines).
The experimental data, taken from \cite{lung} $(\diamond)$,
\cite{bruins} $(\Box)$ and \cite{bartel} $(\ast$), and the calculations
are divided by $\mu_nF_D$.

\vspace{15pt}
\noindent
{\bf Figure 5:}
Comparison between the experimental proton magnetic form factor
$G_M^p$, and the collective form factor
with and without flavor breaking (dashed and solid lines).
The experimental data, taken from \cite{walker}, and the calculations are
divided by $\mu_pF_D$.

\vspace{15pt}
\noindent
{\bf Figure 6:}
Comparison between the experimental ratio of the neutron and proton
magnetic form factors and the calculated ratio using the collective
form factors
with and without flavor breaking (dashed and solid lines).
The experimental values are from \cite{lung,andivahis} $(\diamond)$,
\cite{bruins,walker} $(\Box)$ and \cite{bartel,walker} $(\ast$).

\vspace{15pt}
\noindent
{\bf Figure 7:}
Proton helicity amplitudes for excitation of $N(1520)D_{13}$
(a factor of $+i$ is suppressed).
The calculation with and without flavor breaking are shown by dashed and
solid lines, respectively.
The experimental data are from \cite{burkertdata}.

\vspace{15pt}
\noindent
{\bf Figure 8:}
Same as Figure 7, but for $N(1535)S_{11}$
(a factor of $+i$ is suppressed).

\vspace{15pt}
\noindent
{\bf Figure 9:}
Same as Figure 7, but for $N(1650)S_{11}$
(a factor of $+i$ is suppressed).

\vspace{15pt}
\noindent
{\bf Figure 10:}
Same as Figure 7, but for $N(1680)F_{15}$.

\vspace{15pt}
\noindent
{\bf Figure 11:}
Neutron helicity amplitudes for $N(1680)F_{15}$.

\vspace{15pt}
\noindent
{\bf Figure 12:}
Helicity amplitudes for the excitation of $\Delta(1232)P_{33}$.
The calculation with and without flavor breaking are shown by dashed and
solid lines, respectively. Only the data at $Q^2=0$
(photoproduction) \cite{pdg94} are shown,
since the other experimental results have
not been analyzed in terms of helicity amplitudes.

\vspace{15pt}
\noindent
{\bf Figure 13:}
Helicity-1/2 amplitude for excitation of $\Delta(1620)S_{31}$
(a factor of $+i$ is suppressed).
The calculation with and without flavor breaking are shown by dashed and
solid lines, respectively.
The experimental data are from \cite{pdg94} and \cite{burkertdata}.

\vspace{15pt}
\noindent
{\bf Figure 14:}
Same as Figure 13, but for $\Delta(1700)D_{33}$
(a factor of $+i$ is suppressed).

\vspace{15pt}
\noindent
{\bf Figure 15:}
Same as Figure 13, but for helicity-3/2 amplitude of $\Delta(1700)D_{33}$
(a factor of $+i$ is suppressed).

\vspace{15pt}
\noindent
{\bf Figure 16:}
Helicity asymmetry for $\Delta(1232)P_{33}$.
The experimental data are from \cite{pdg94}.

\vspace{15pt}
\noindent
{\bf Figure 17:}
Proton helicity asymmetry for $N(1520)D_{13}$.
The experimental data are from \cite{burkertdata}.

\vspace{15pt}
\noindent
{\bf Figure 18:}
Same as Figure 17, but for $N(1680)F_{15}$.

\vspace{15pt}
\noindent
{\bf Figure 19:}
Proton helicity amplitude for excitation of $N(1440)P_{11}$.
The calculation with and without flavor breaking are shown by dashed and
solid lines, respectively. The curves are labelled by the value
of $\chi$ (see Table~\ref{tamp}).
The experimental data are from \cite{burkertdata}.

\vspace{15pt}
\noindent
{\bf Figure 20:}
Effect of hadron swelling for excitation of $\Delta(1232)P_{33}$.
The curves are labelled by the value of the stretching parameter $\xi$
of Eq.~(\ref{stretch}).

\vspace{15pt}
\noindent
{\bf Figure 21:}
Same as Figure 20, but for $N(1520)D_{13}$
(a factor of $+i$ is suppressed).

\vspace{15pt}
\noindent
{\bf Figure 22:}
Same as Figure 20, but for $N(1680)F_{15}$.

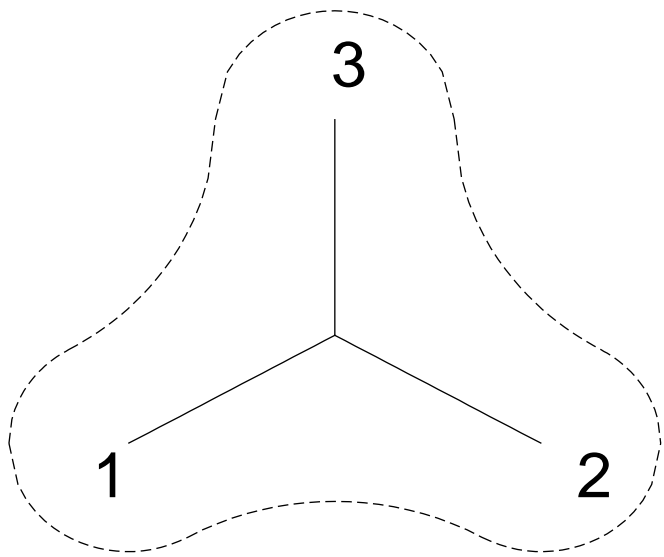
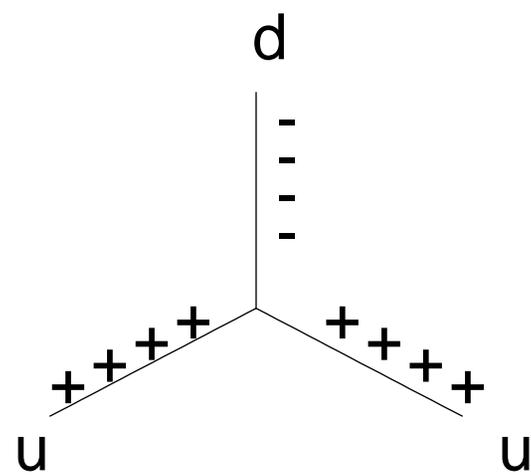
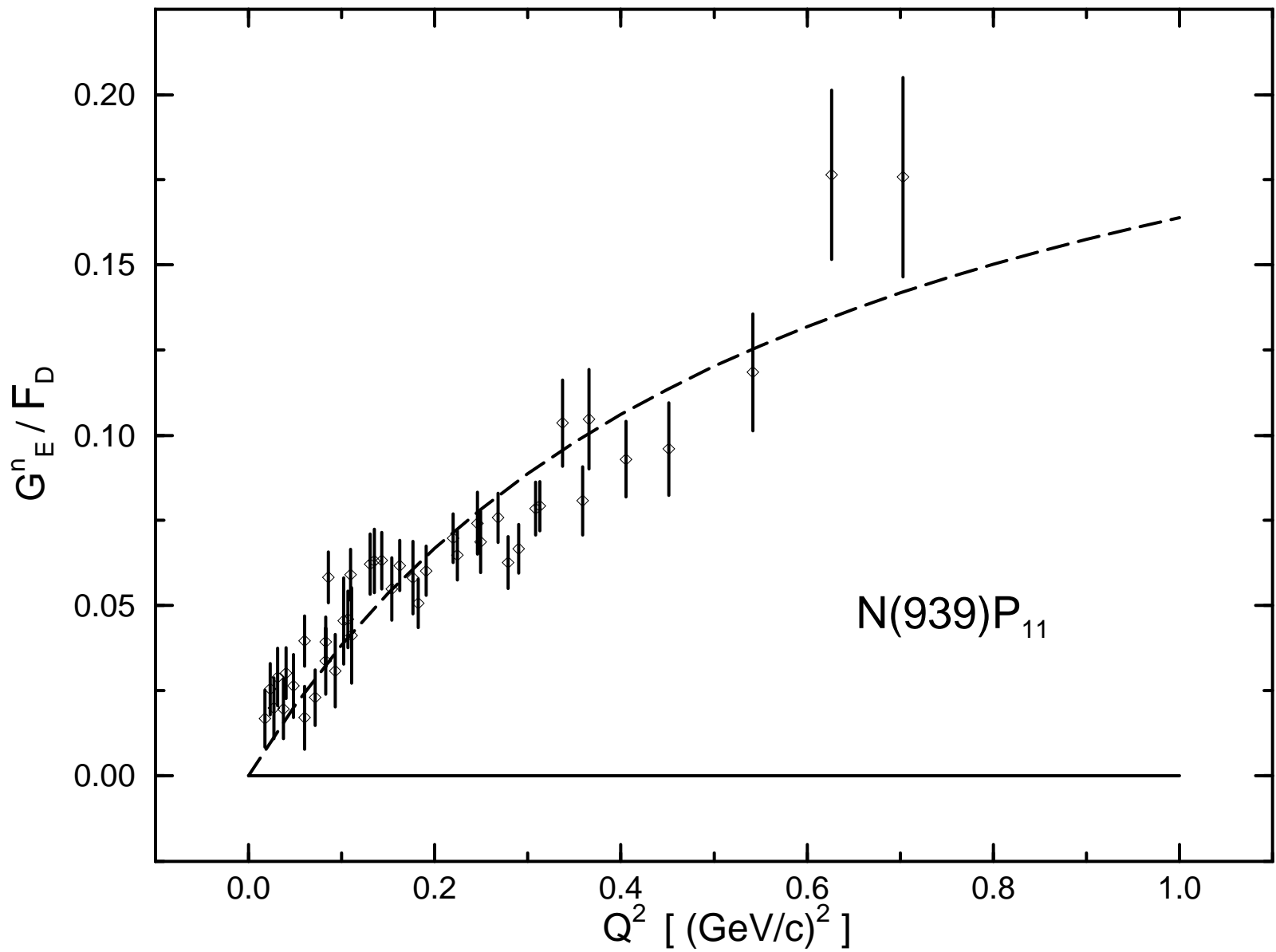
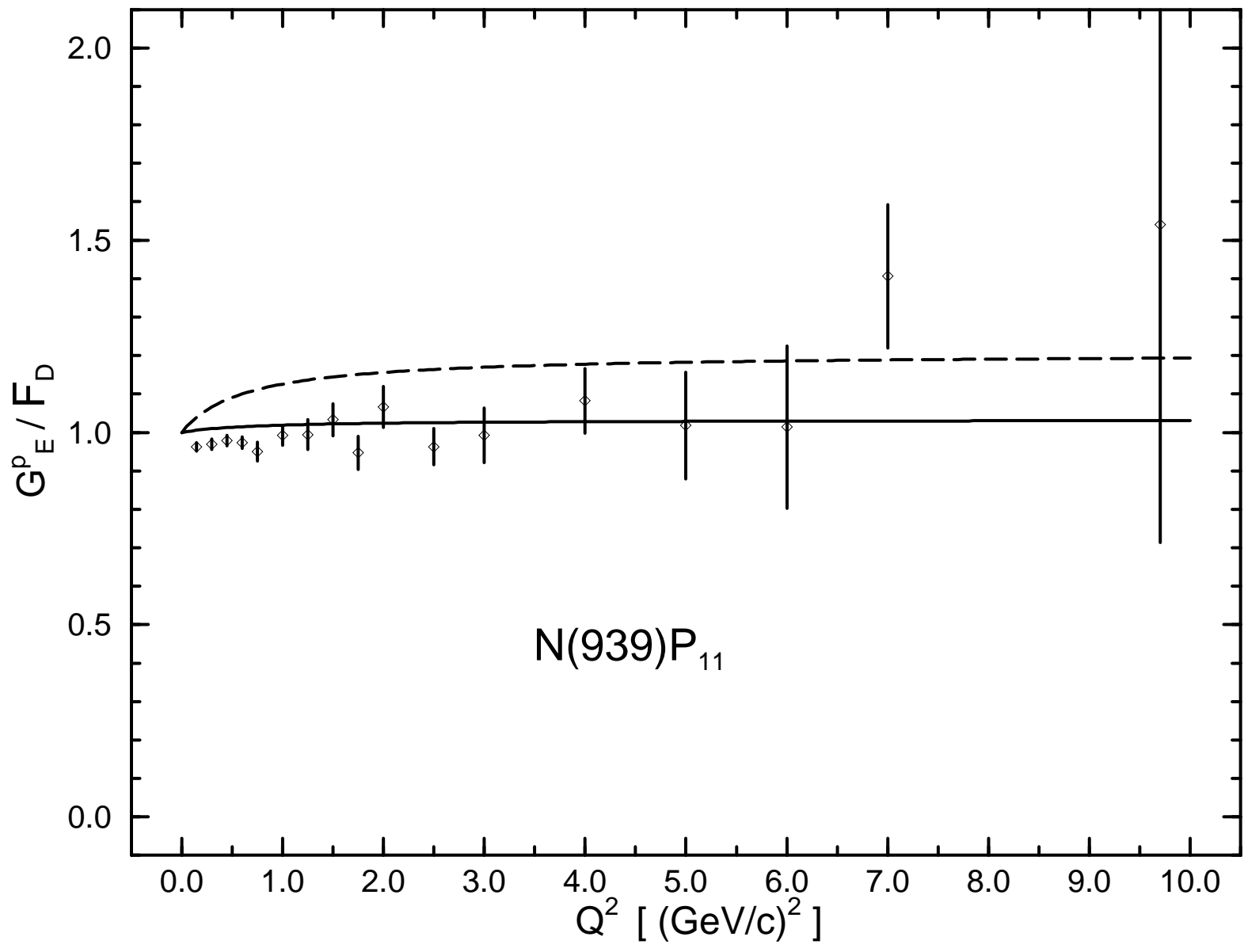
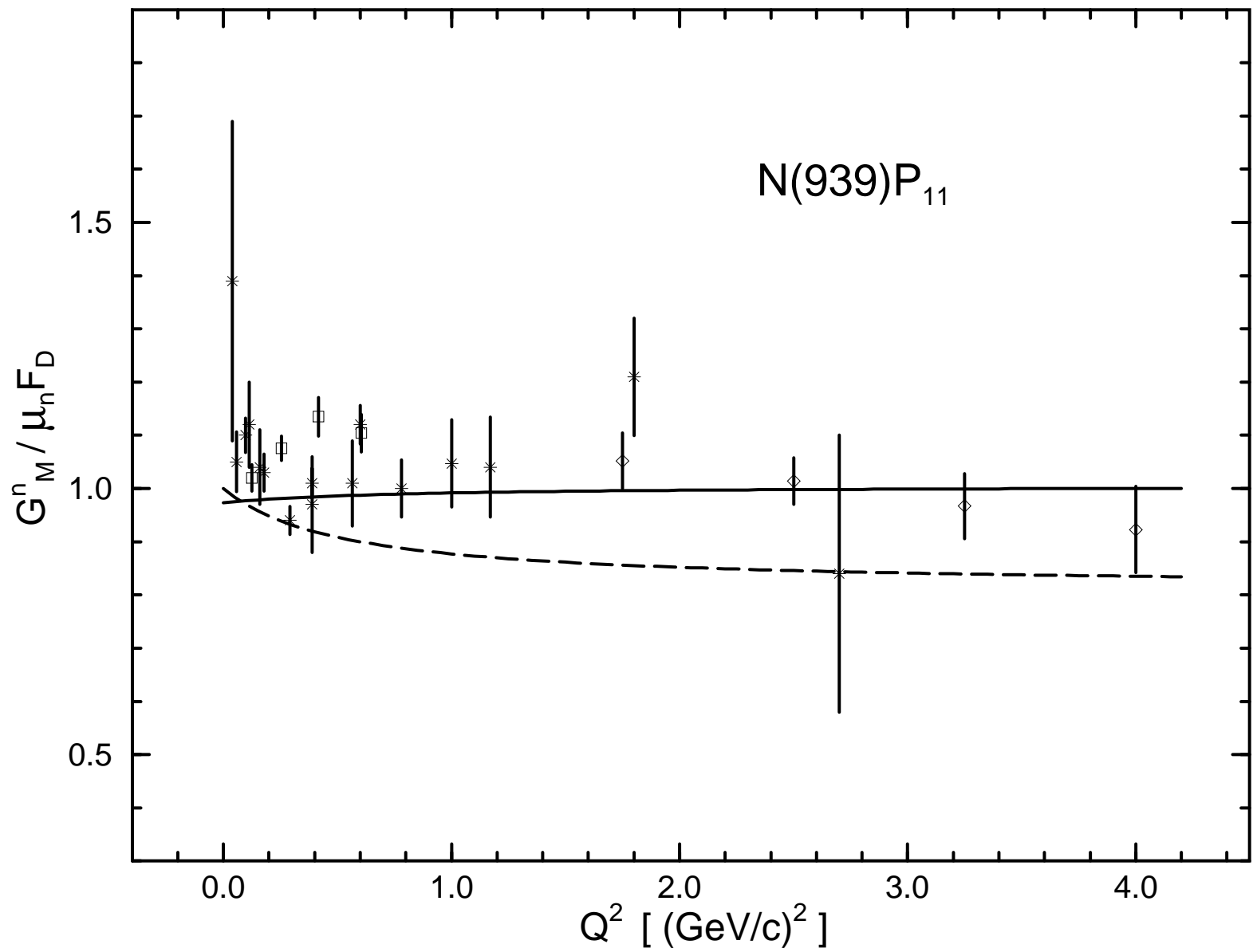
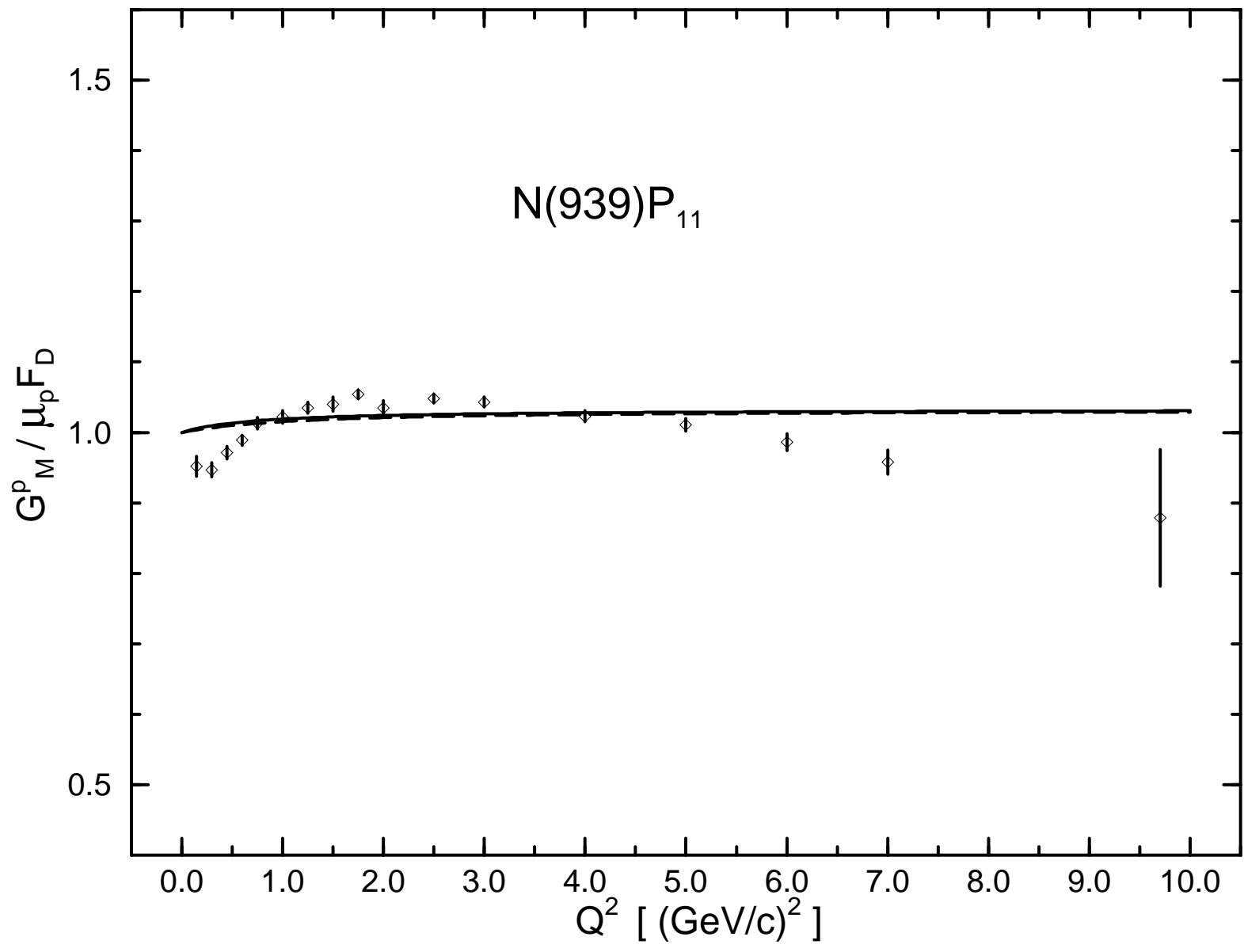
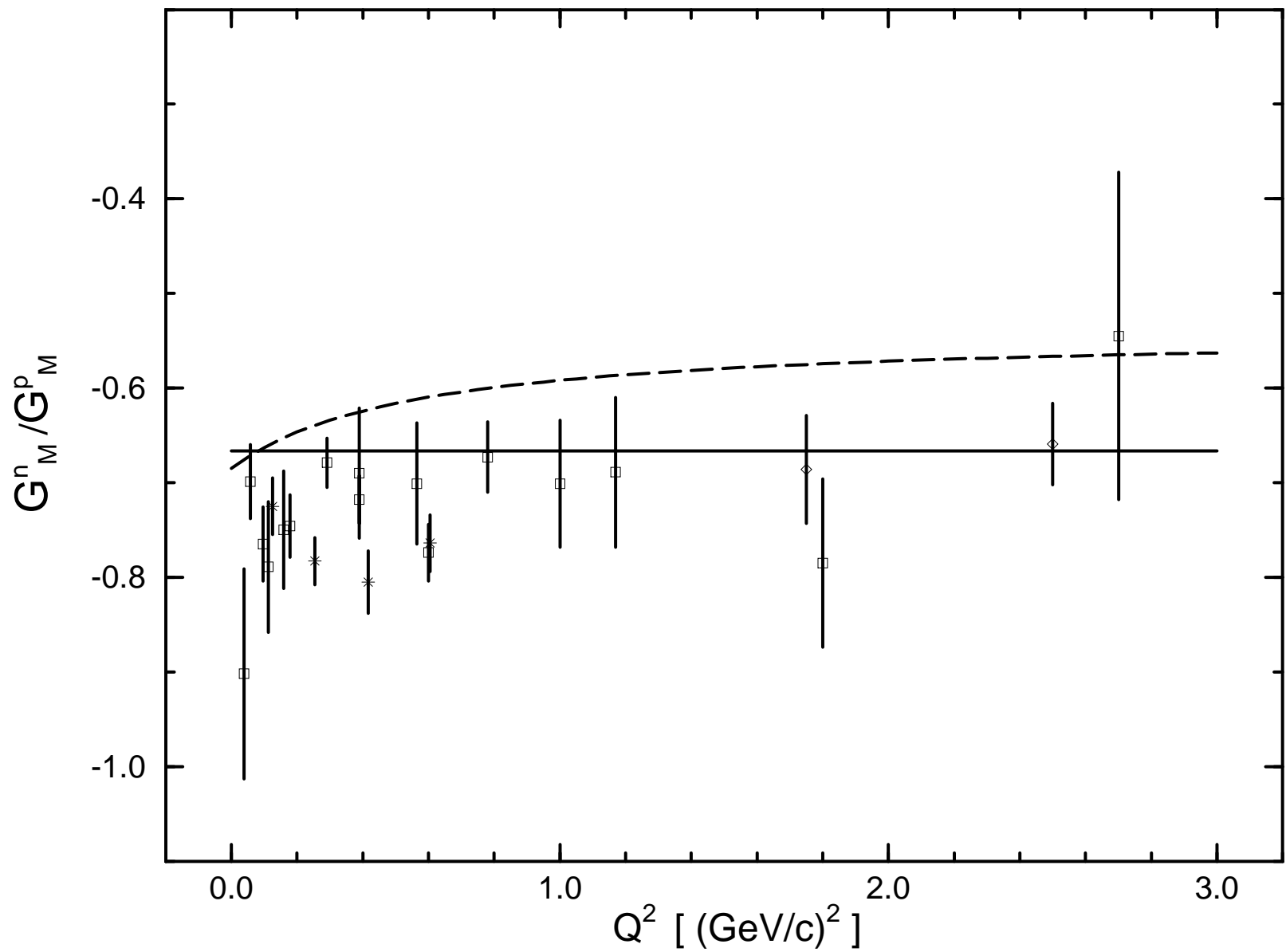
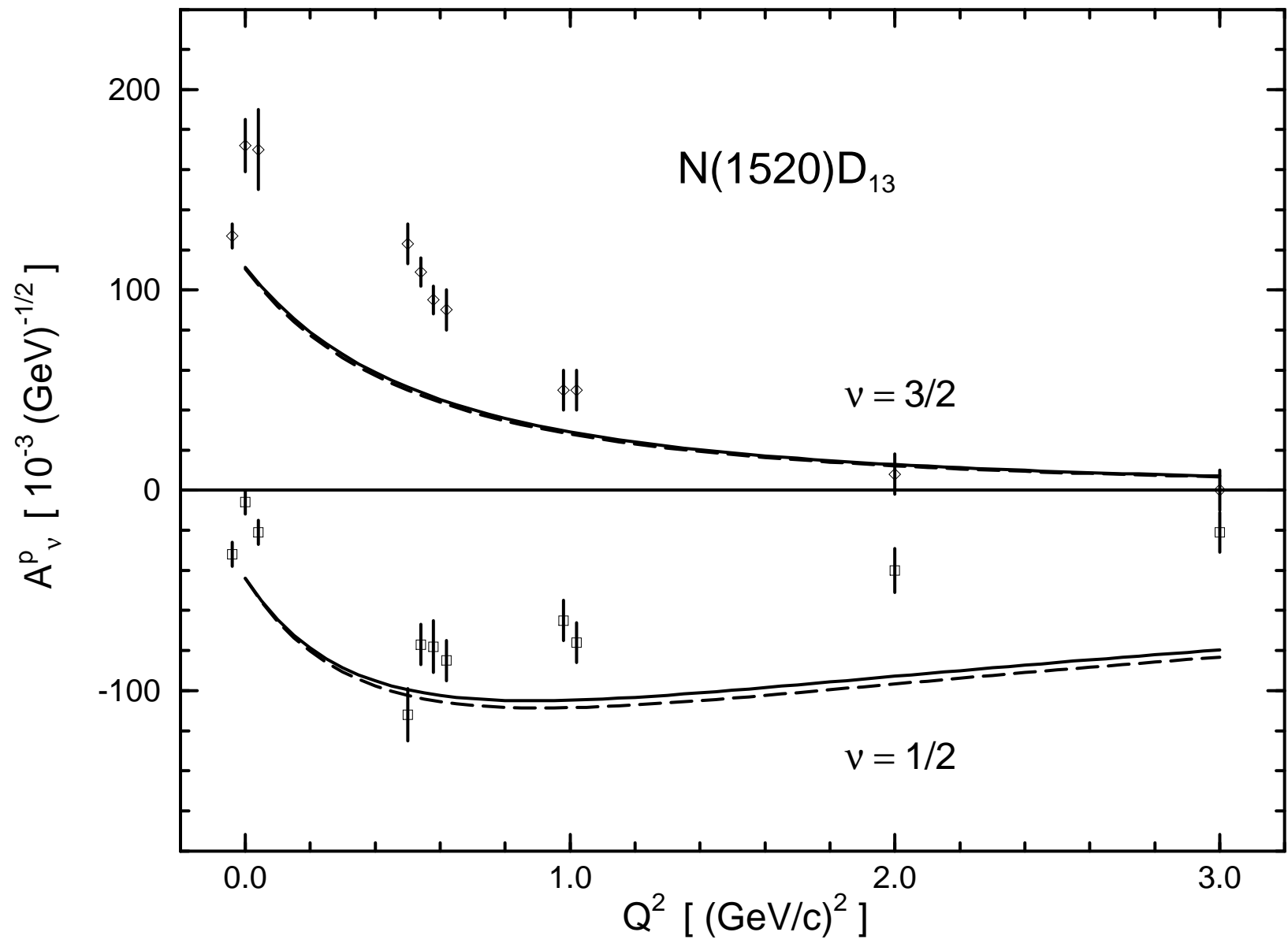
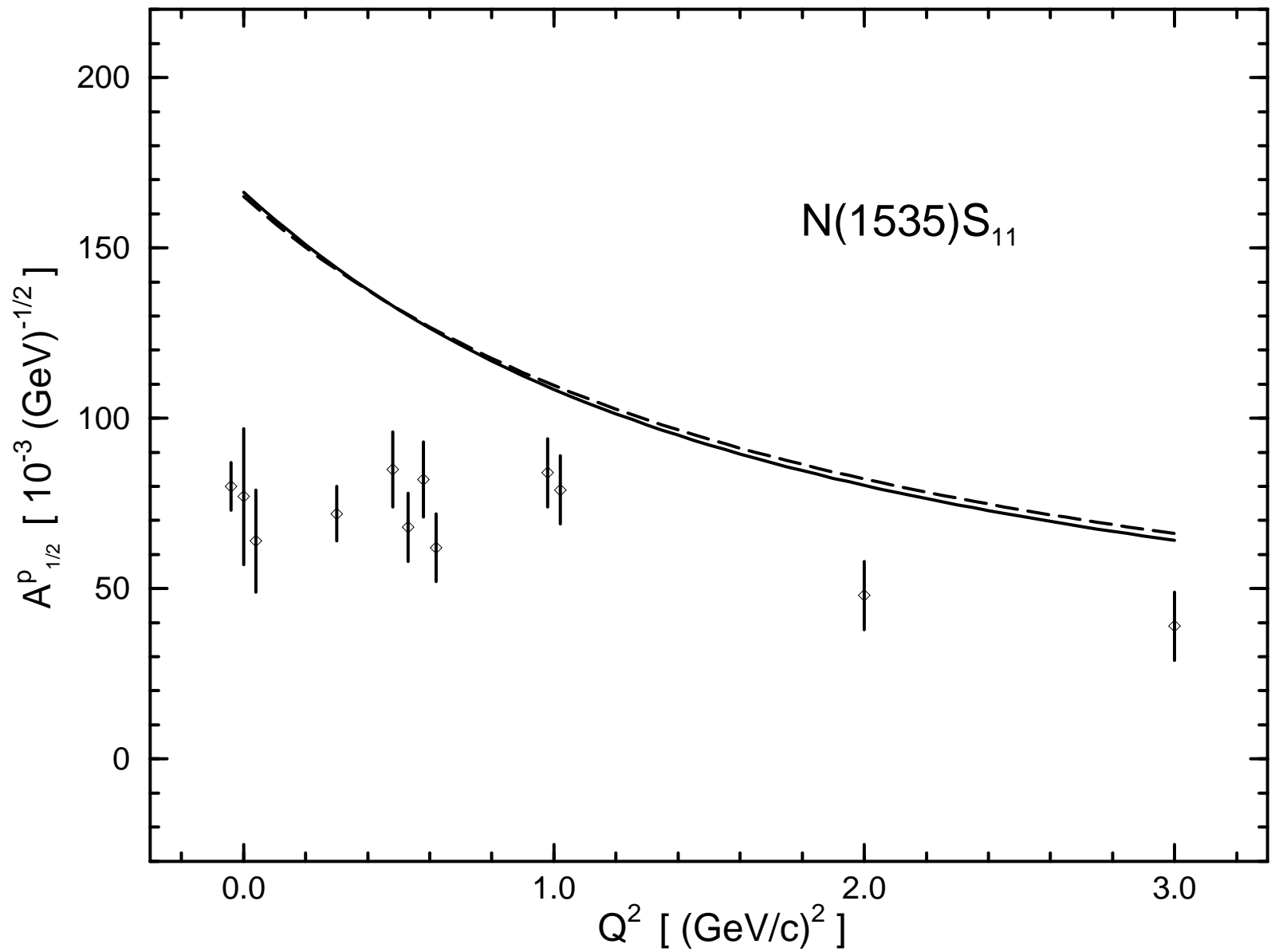
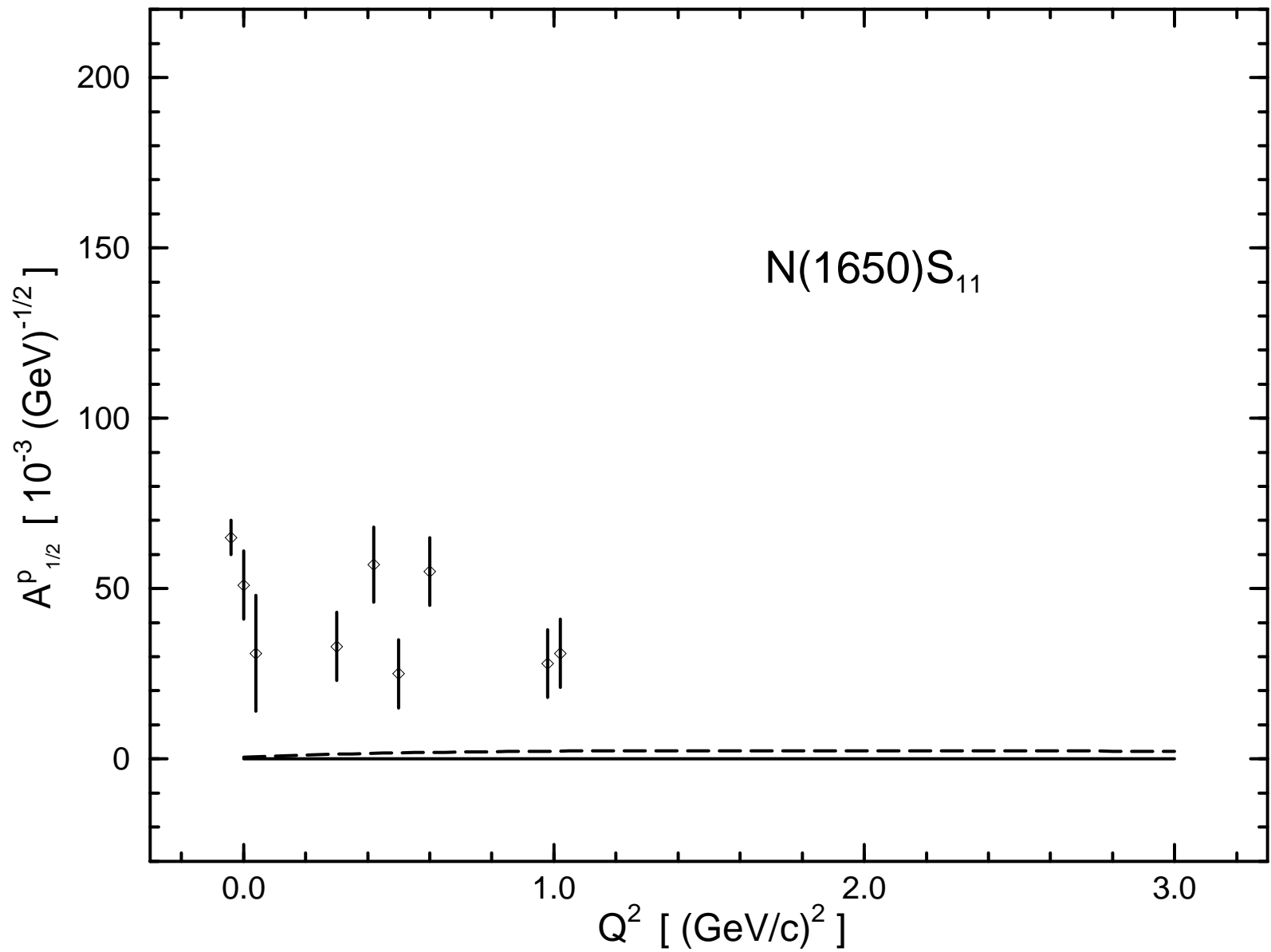
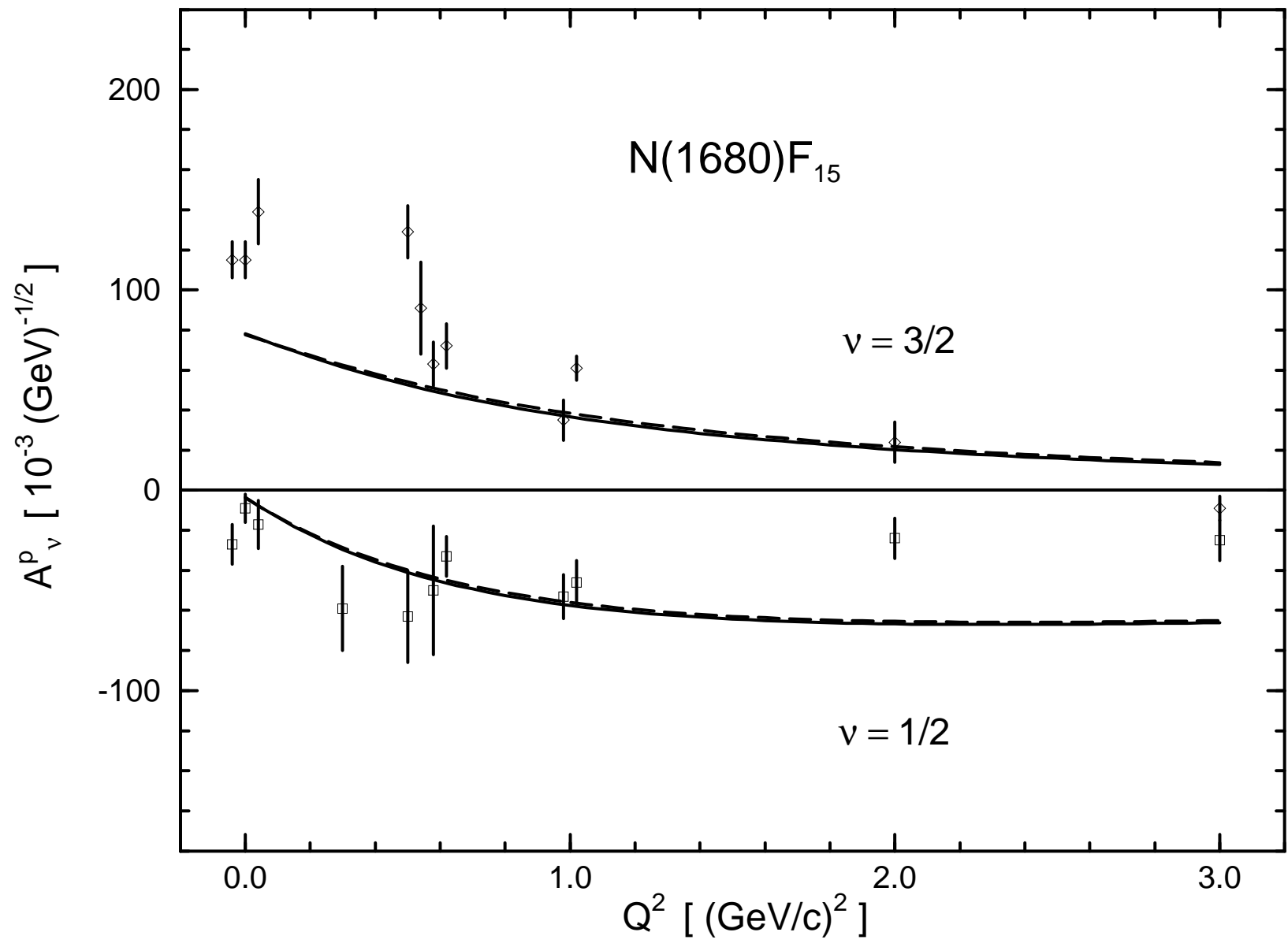
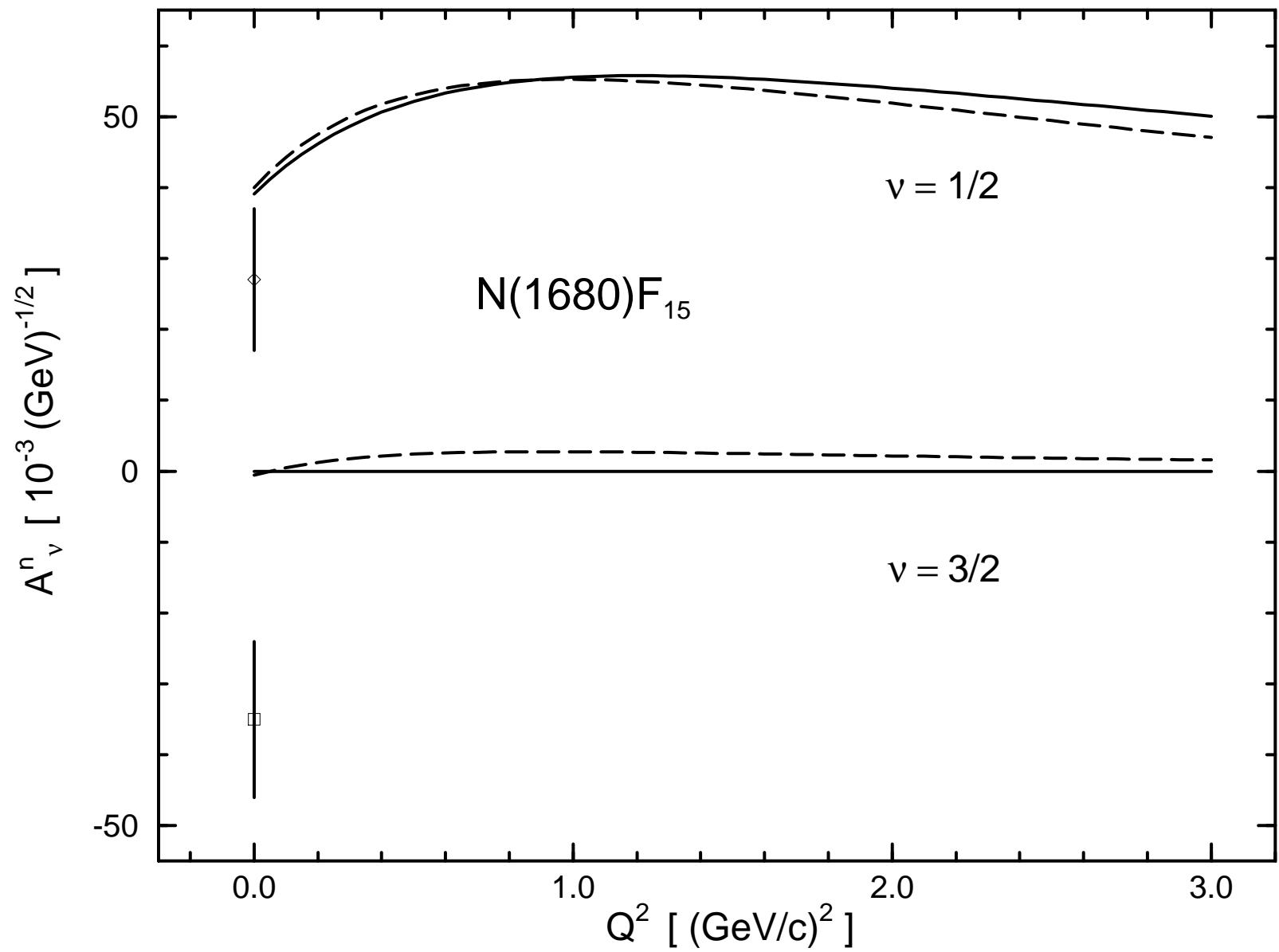
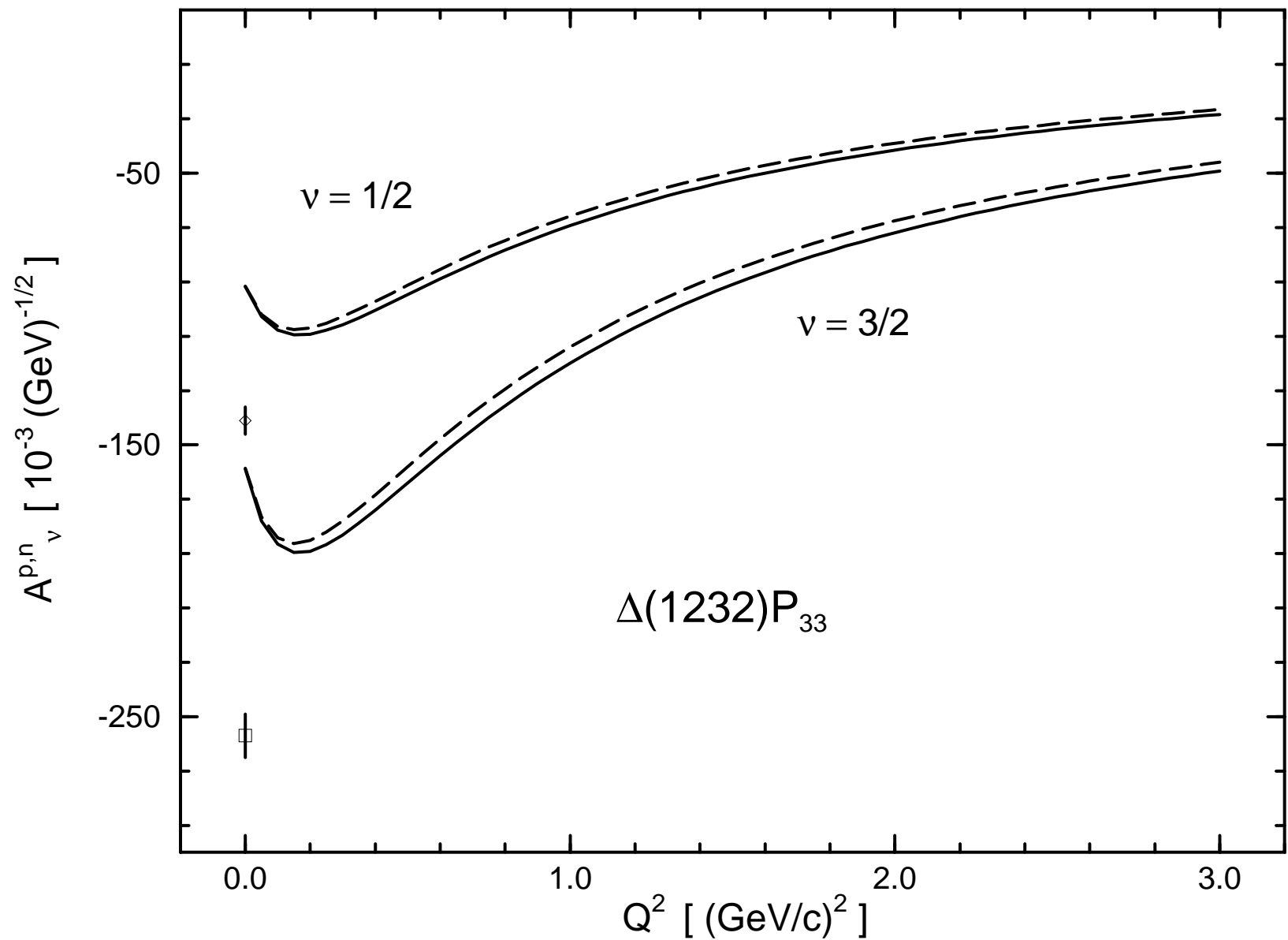
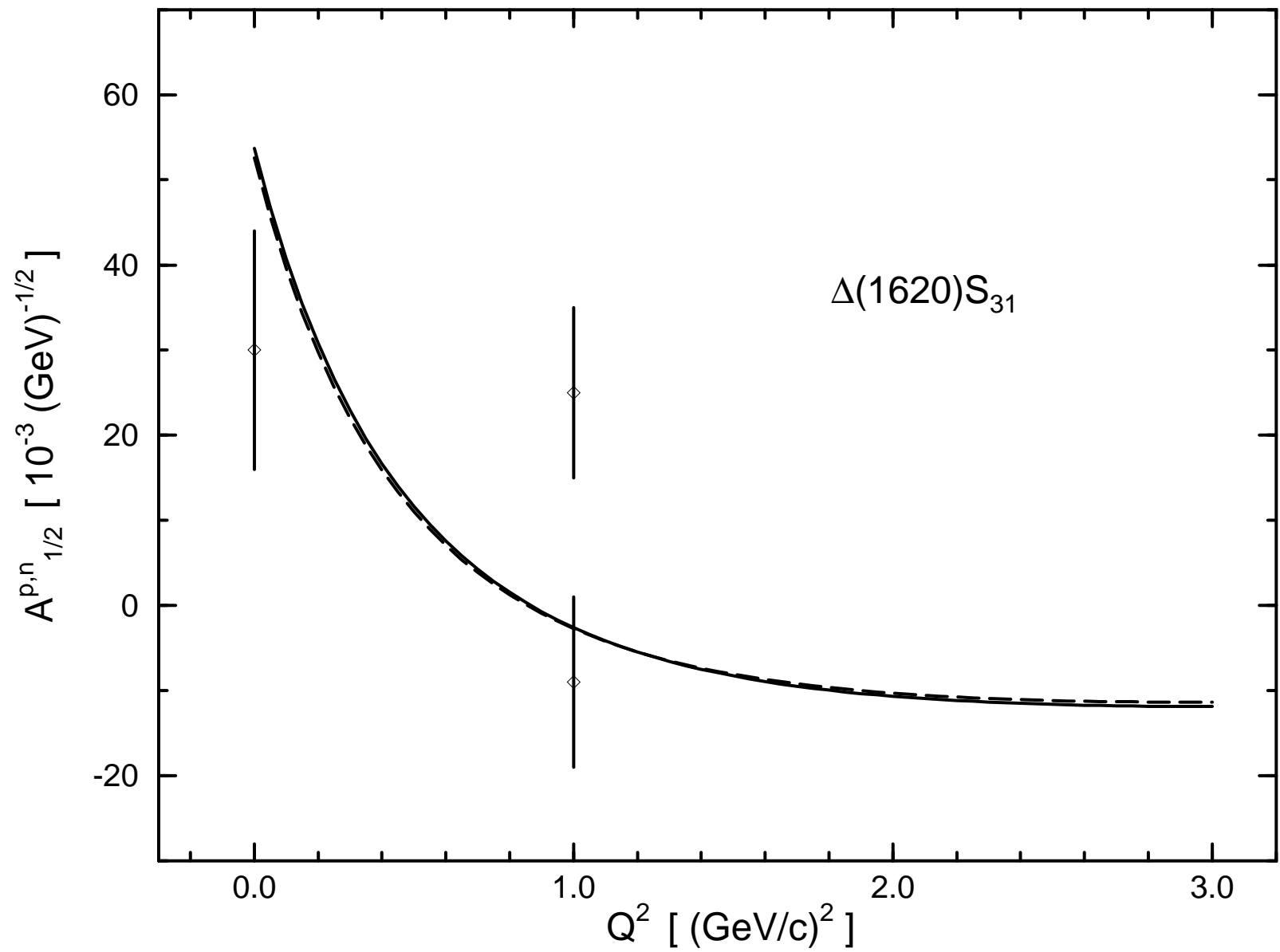
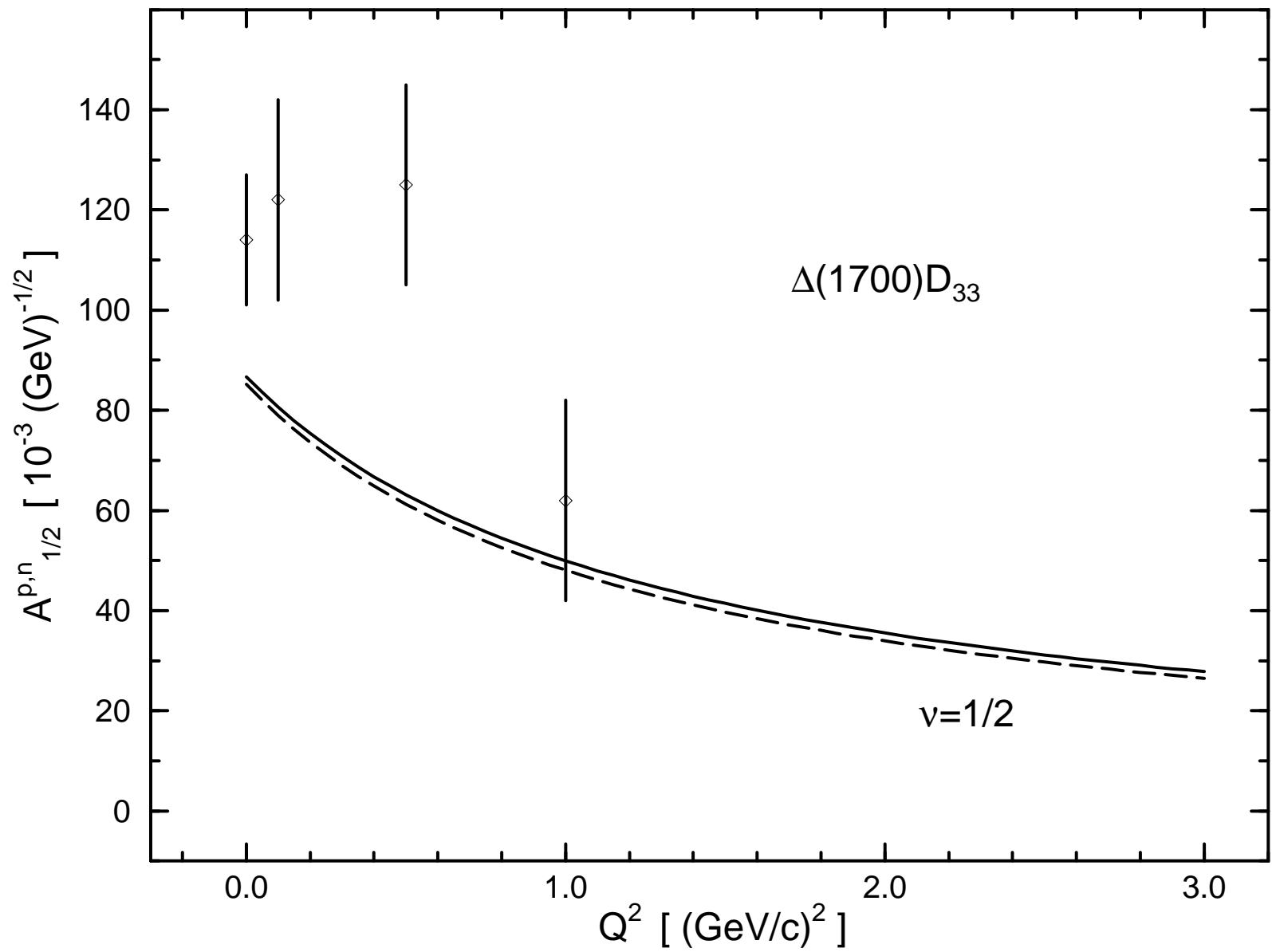
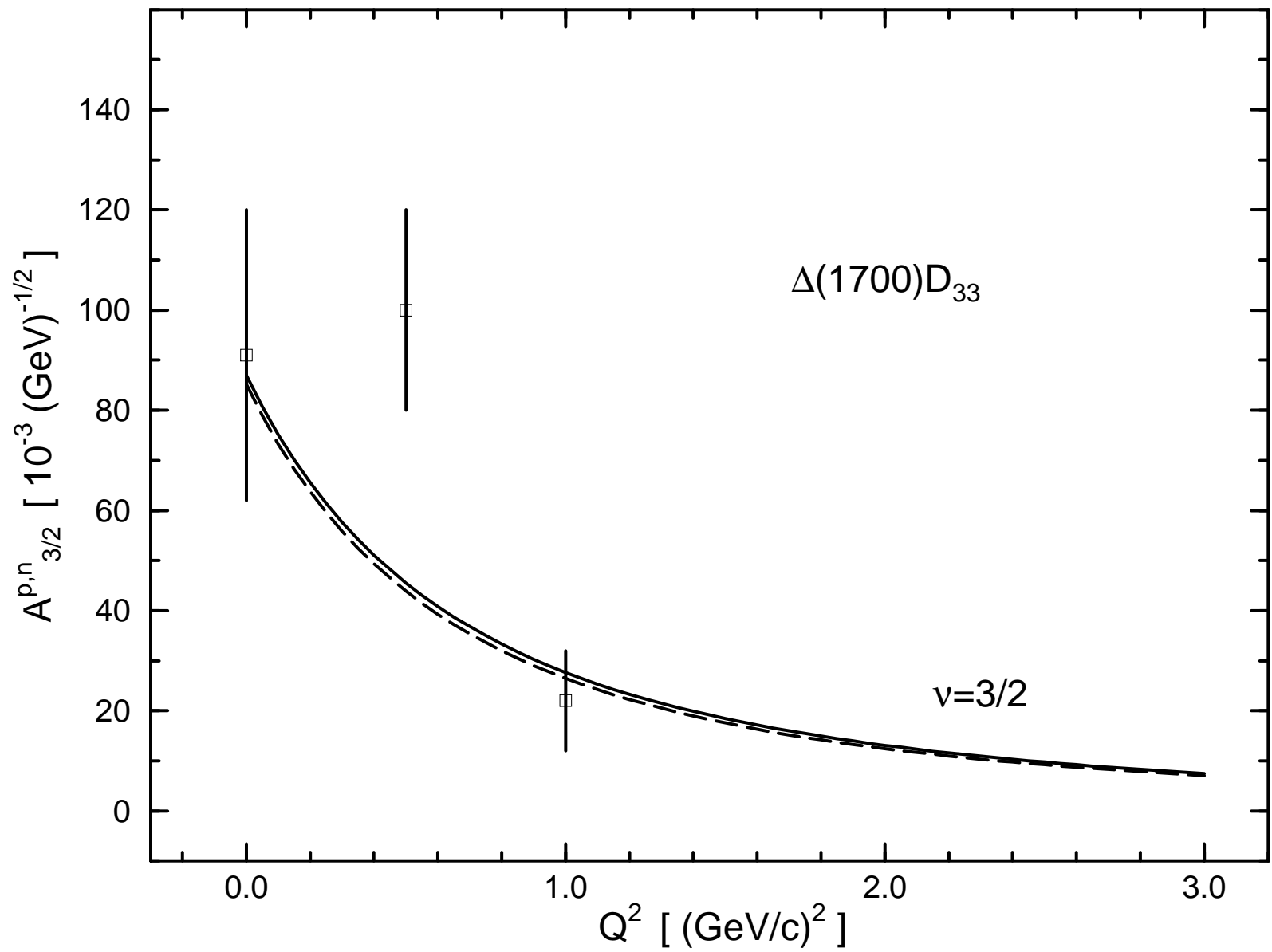
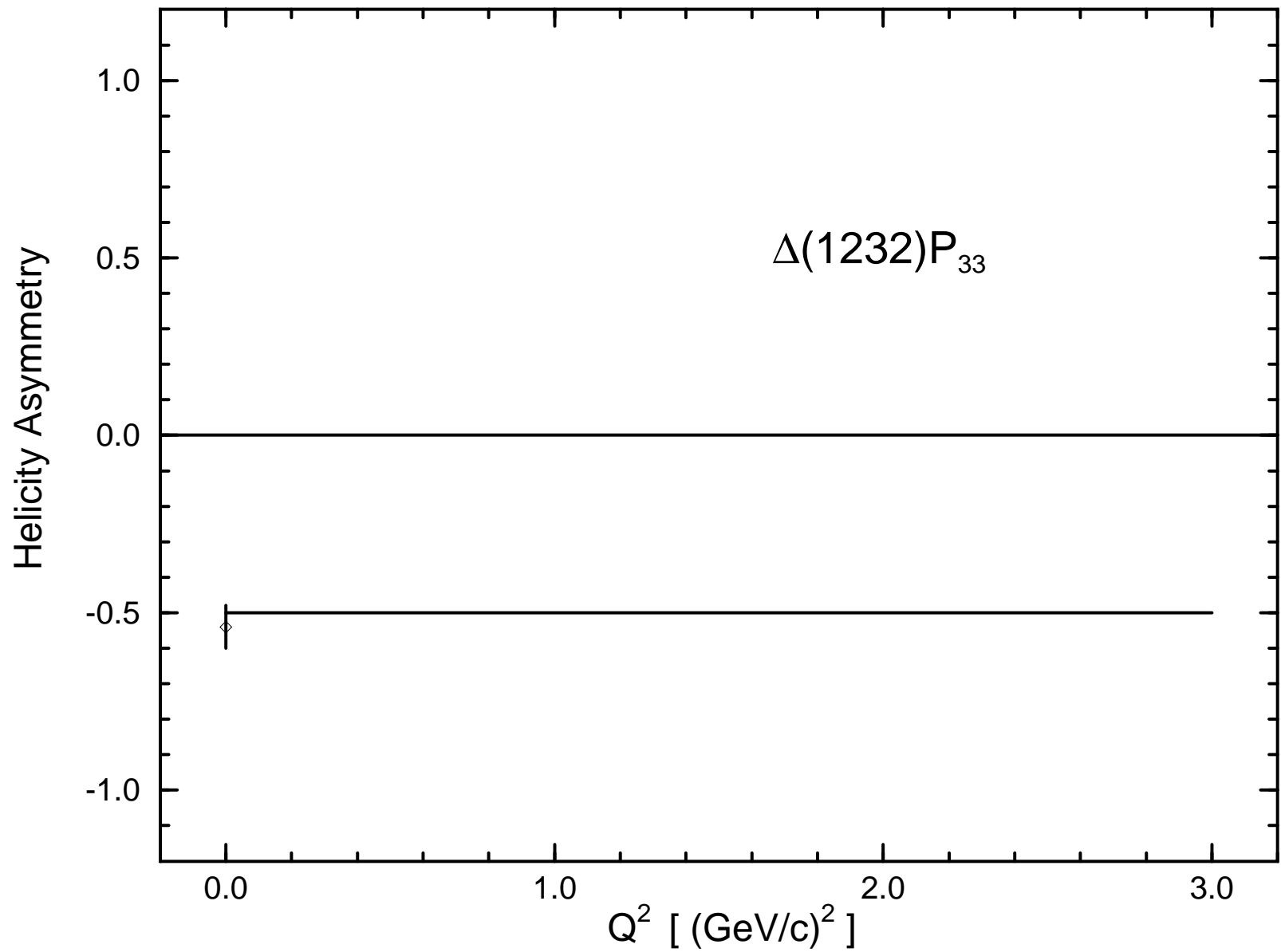
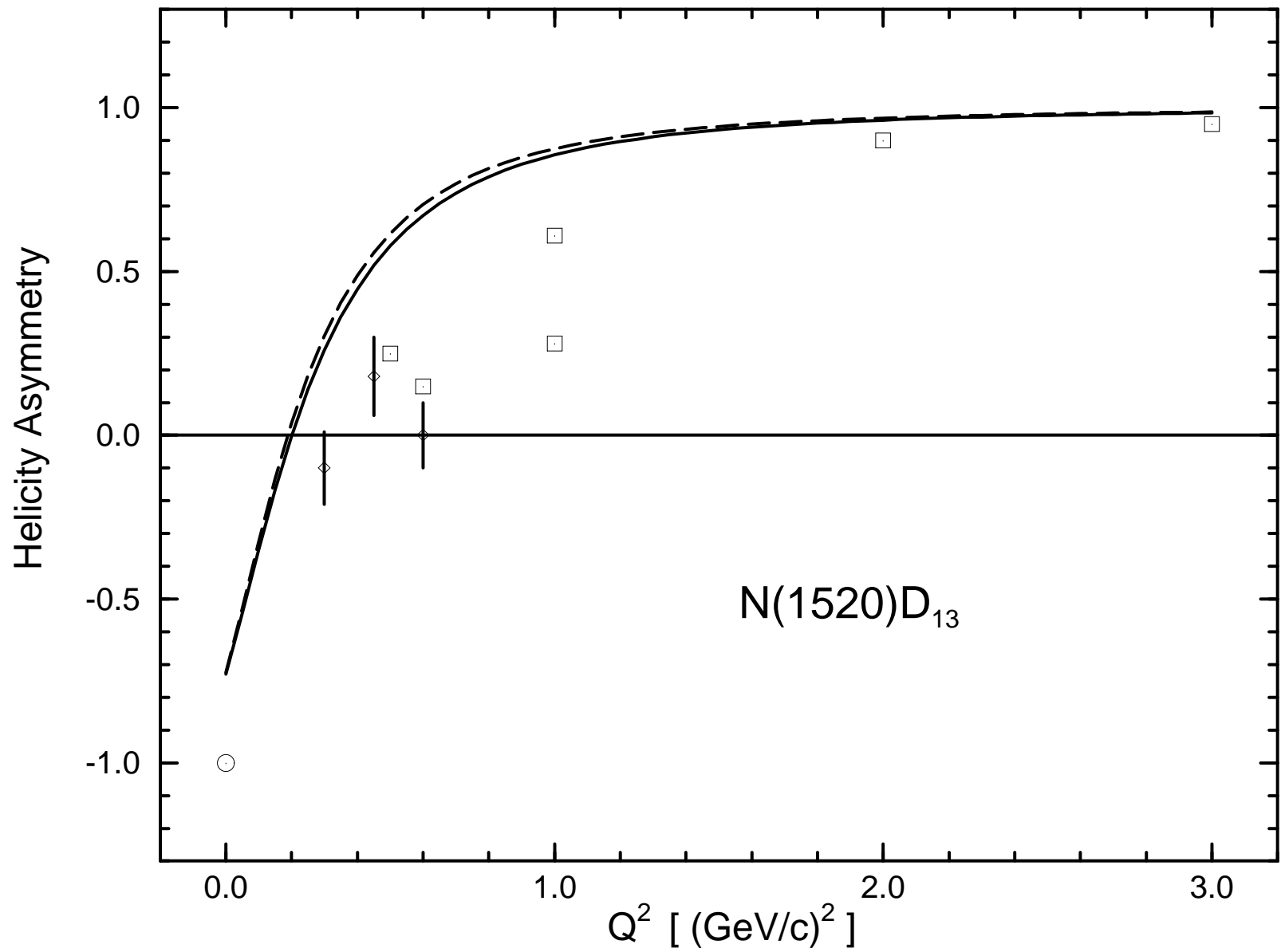
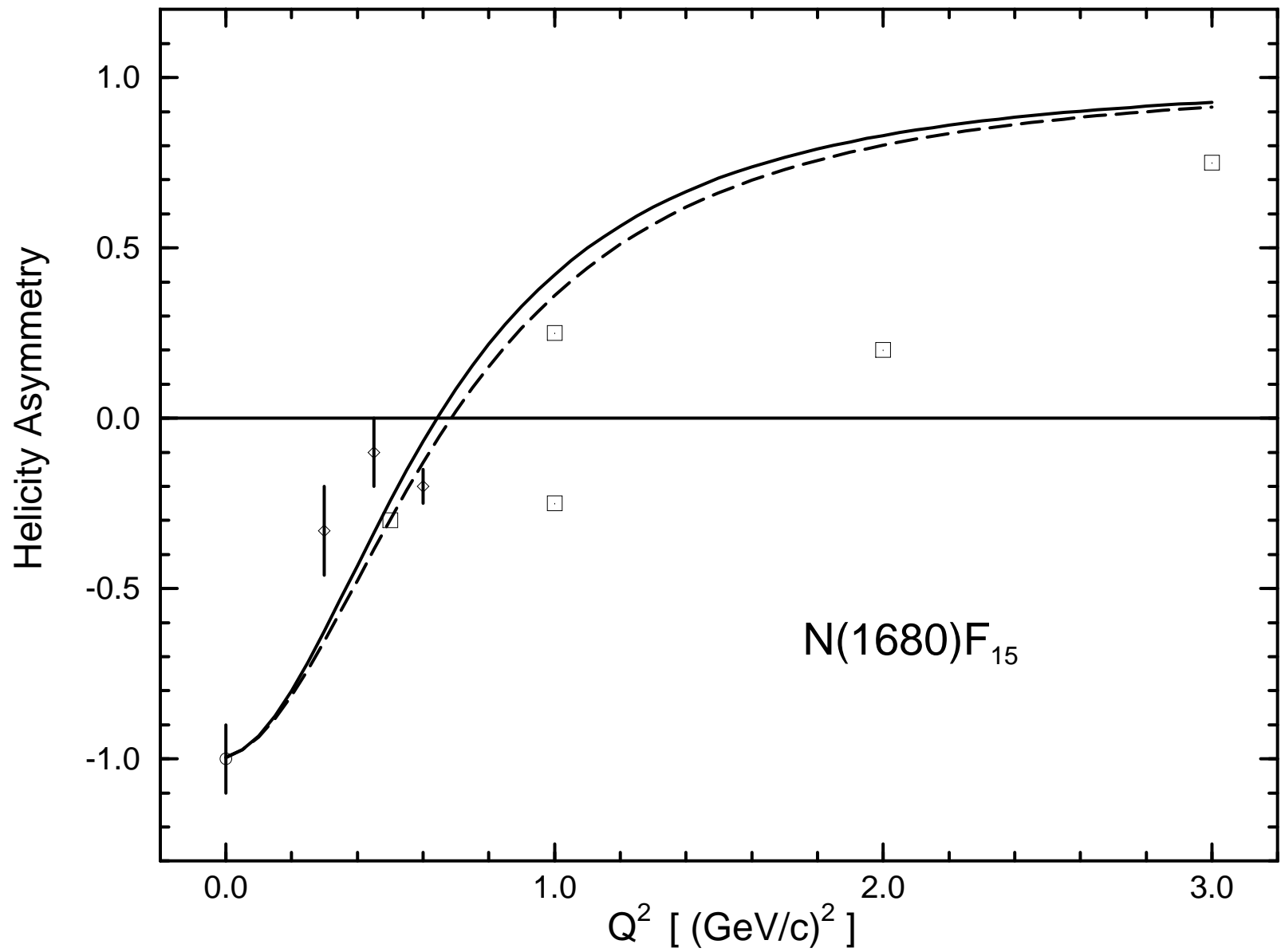
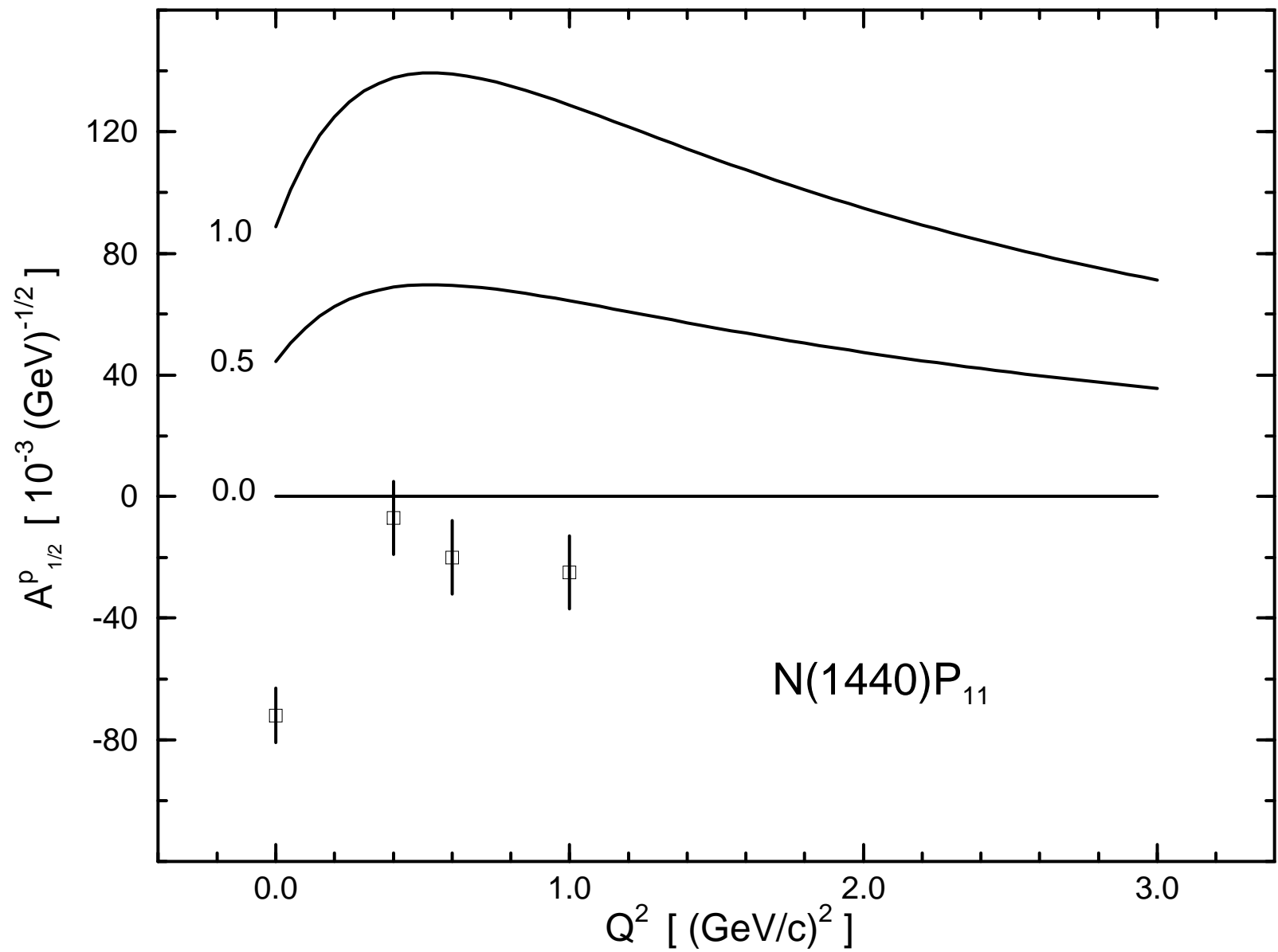
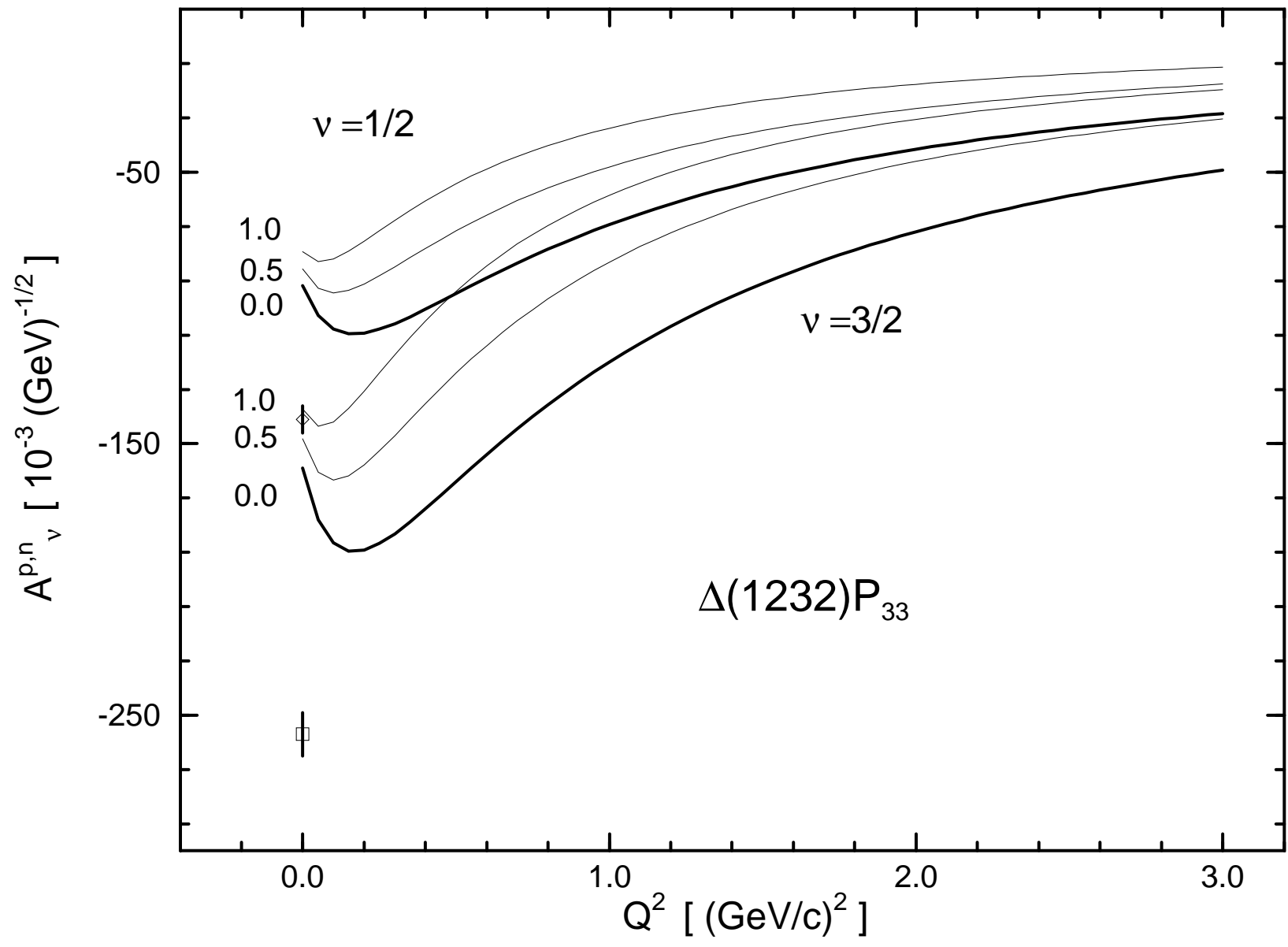
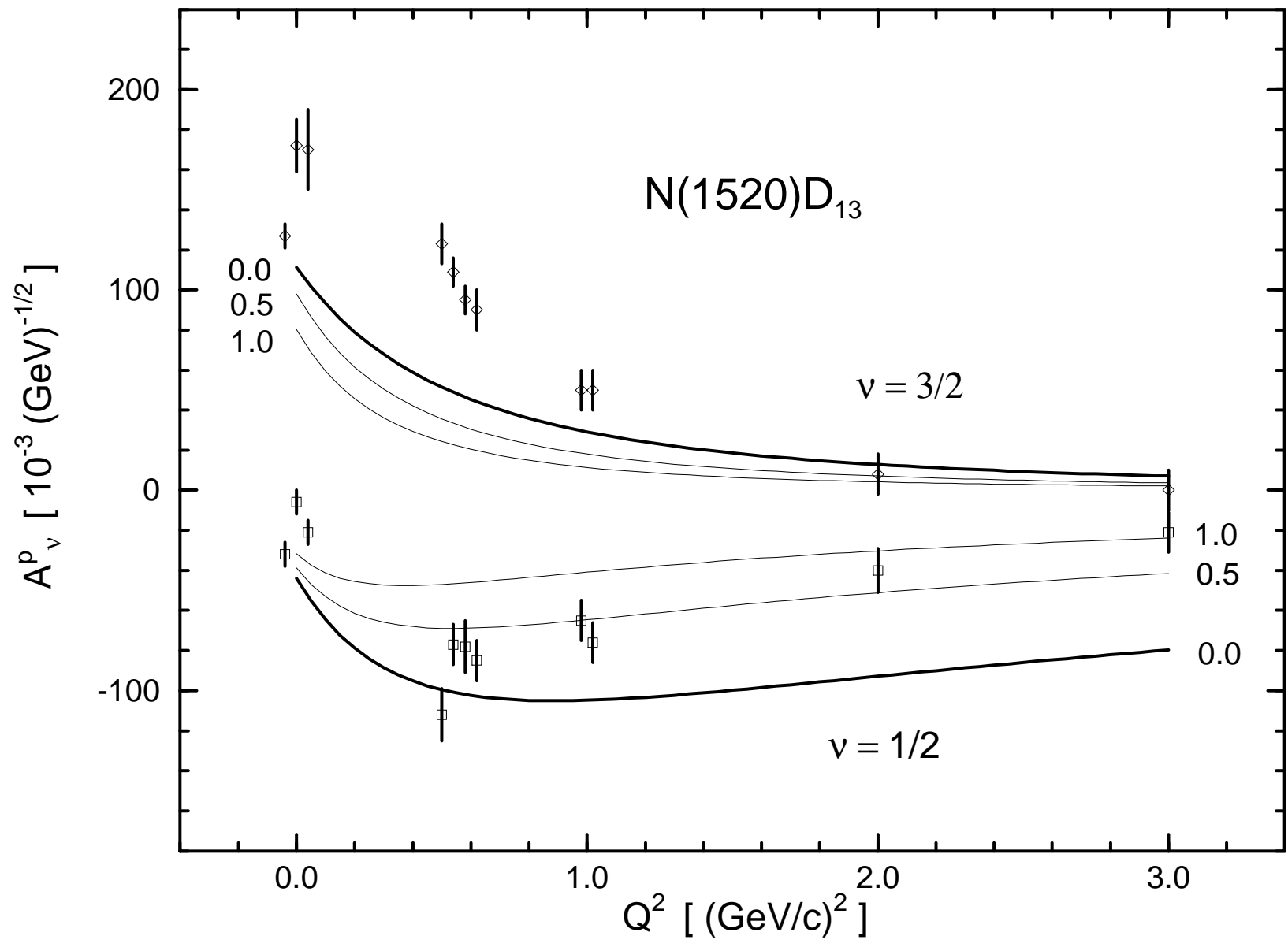
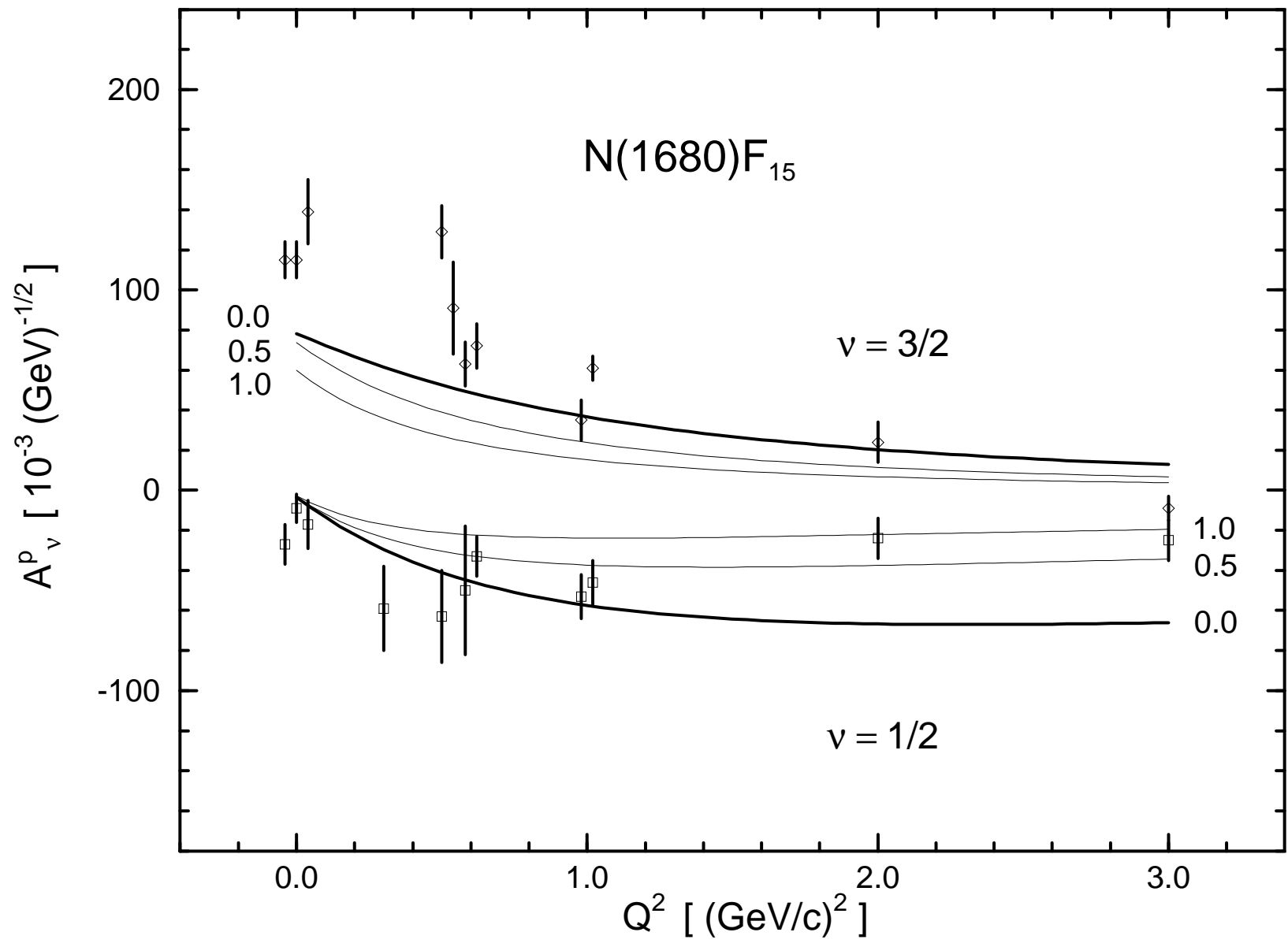
\end{document}